\documentstyle[12pt]{article}
\newlength{\abstractwidth}
\begin{document}
\thispagestyle{empty}
\pagestyle{empty}
\renewcommand{\thefootnote}{\fnsymbol{footnote}}
\renewcommand{\title}[1]{\begin{center}\large\bf #1\end{center}\par}
\renewcommand{\author}[1]{\vspace{2ex}{\normalsize\begin{center}
\setlength{\baselineskip}{3ex}#1\par\end{center}}}
\renewcommand{\thanks}[1]{\footnote{#1}} 
\renewcommand{\abstract}[1]{\vspace{2ex}\normalsize\begin{center}
\centerline{\bf Abstract}\par\vspace{2ex}\parbox{\abstractwidth}{{\small #1
\setlength{\baselineskip}{2.5ex}\par}}
\end{center}}
\newcommand{\starttext}{\normalsize
\pagestyle{plain}
\setlength{\baselineskip}{14pt}\par
\setcounter{footnote}{0}
\renewcommand{\thefootnote}{\arabic{footnote}}}


\def\ang{\,{\rm\AA}}
\def\flux{\,{\rm erg\,cm^{-2}\,arcsec^{-2}\,\AA^{-1}\,s^{-1}}}
\def\GeV{\,{\rm GeV}}
\def\TeV{\,{\rm TeV}}
\def\gev{\,{\rm GeV}}
\def\keV{\,{\rm keV}}
\def\MeV{\,{\rm MeV}}
\def\sec{\,{\rm sec}}
\def\Gyr{\,{\rm Gyr}}
\def\yr{\,{\rm yr}}
\def\rcm{\,{\rm cm}}
\def\pc{\,{\rm pc}}
\def\kpc{\,{\rm kpc}}
\def\Mpc{\,{\rm Mpc}}
\def\mpc{\,{\rm Mpc}}
\def\eV{{\,\rm eV}}
\def\ev{{\,\rm eV}}
\def\erg{{\,\rm erg}}
\def\cmm2{{\,\rm cm^{-2}}}
\def\cm2{{\,{\rm cm}^2}}
\def\cmm3{{\,{\rm cm}^{-3}}}
\def\gcmm3{{\,{\rm g\,cm^{-3}}}}
\def\kms{\,{\rm km\,s^{-1}}}
\def\HO{{100h\,{\rm km\,sec^{-1}\,Mpc^{-1}}}}
\def\mpl{{m_{\rm Pl}}}
\def\mpp{{m_{\rm Pl,0}}}
\def\trh{T_{\rm RH}}
\def\g{\tilde g}
\def\R{{\cal R}}
\def\km{\rm \,km}
\def\yrs{\rm \,yrs}
\def\trh{T_{\rm RH}}

\def\baselinestretch{1.4}
\def\VEV#1{\left\langle #1\right\rangle}
\def\la{\mathrel{\mathpalette\fun <}}
\def\ga{\mathrel{\mathpalette\fun >}}
\def\fun#1#2{\lower3.6pt\vbox{\baselineskip0pt\lineskip.9pt
  \ialign{$\mathsurround=0pt#1\hfil##\hfil$\crcr#2\crcr\sim\crcr}}}

\begin{flushright}
FERMILAB-Conf-92/313-A
\end{flushright}

\title{INFLATION AFTER COBE\\
Lectures on Inflationary Cosmology\thanks{Summary of lectures
given at TASI-92 (Boulder, CO, June 1992), Cargese Summer School
on Quantitative Particle Physics (Cargese, Corsica, July 1992), and
Varenna Summer School on Galaxy Formation (Varenna, Italy,
July 1992).}
}

\author{Michael S. Turner\\
{\small\it Departments of Physics and Astronomy \& Astrophysics,\\
Enrico Fermi Institute, The University of Chicago, Chicago, IL~~60637-1433}\\
\vspace{.1in}
{\it NASA/Fermilab Astrophysics Center,\\
Fermi National Accelerator Laboratory, Batavia, IL~~60510-0500}
}

\date{}
\abstract{\small
In these lectures I review the standard
hot big-bang cosmology, emphasizing its successes, its shortcomings,
and its major challenge---a detailed
understanding of the formation of structure in the
Universe.  I then discuss the motivations
for---and the fundamentals
of---inflationary cosmology, particularly emphasizing the
quantum origin of metric (density and gravity-wave) perturbations.
Inflation addresses the shortcomings of the standard cosmology
and provides the ``initial data'' for structure formation.
I conclude by addressing the implications of inflation for structure formation,
evaluating the various cold dark matter models
in the light of the recent detection of temperature
anisotropies in the cosmic background radiation by COBE.  In the near term,
the study of structure formation offers a powerful probe
of inflation, as well as specific inflationary models.}

\starttext

\section{Hot Big Bang:  Successes and Challenges}
\subsection{Successes}

The hot big-bang model, more properly the Friedmann-Robertson-Walker (FRW)
cosmology or standard cosmology, is spectacularly successful:
In short, it provides a reliable and tested accounting
of the history of the Universe from about $0.01\sec$
after the bang until today, some 15 billion years later.
The primary pieces of evidence that support the model
are:  (1) The expansion of the Universe;
(2) The cosmic background radiation; and (3) The
primordial abundances of the light elements D, $^3$He,
$^4$He, and $^7$Li \cite{standard}.

\subsubsection{The expansion}

Although the precise value of the Hubble constant
is not known to better than a factor of two, $H_0 =100h\,\km
\sec^{-1}\Mpc^{-1}$ with $h=0.4-1$, there is little doubt
that the expansion obeys the ``Hubble law'' out to red shifts
approaching unity \cite{h50,mould};
see Fig.~1.  As is well appreciated,
the fundamental difficulty in determining the Hubble constant
is the calibration of the cosmic-distance scale as ``standard
candles'' are required \cite{distance1,distance2}.

\begin{figure}
\vspace{4.0in}
\caption[hubble]{Hubble diagram (from \cite{mould}).
The deviation from a linear relationship around $40\Mpc$
is due to peculiar velocities.}
\end{figure}

The Hubble law allows one to
infer the distance to an object from its red shift $z$:
$d = zH_0^{-1}\simeq 3000z\,h^{-1}\Mpc$ (for $z\ll 1$,
the galaxy's recessional velocity $v\simeq zc$), and hence ``maps
of the Universe'' constructed from galaxy positions and red shifts
are referred to as red-shift surveys.  Ordinary galaxies and clusters of
galaxies are seen out to red shifts of order unity; more
unusual and rarer objects, such as radio galaxies and quasars,
are seen out to red shifts of almost five (the current
record holder is a quasar with red shift 4.9).  Thus, we can
probe the Universe with visible light to within a few billion years
of the big bang.

\subsubsection{The cosmic background radiation}

The spectrum of the cosmic background radiation (CBR) is consistent
that of a black body at temperature 2.73 K over more than three
decades in wavelength ($\lambda \sim 0.03\rcm - 100\rcm$); see
Fig.~2.  The most accurate measurement of the temperature
and spectrum is that by the FIRAS instrument on the
COBE satellite which determined its temperature to be
$2.726\pm 0.01\,$K \cite{FIRAS}.  It is difficult to
come up with a process other
than an early hot and dense phase in the history
of the Universe that would lead to such a precise
black body \cite{dnsnature}.  According to the standard cosmology,
the surface of last scattering for the CBR is
the Universe at a red shift of about $1100$ and
an age of about $180,000\,(\Omega_0 h^2)^{-1/2}\yrs$.
It is possible that the Universe became ionized
again after this epoch, or due to energy injection
never recombined; in this case the last-scattering surface
is even ``closer,'' $z_{\rm LSS} \simeq 10[\Omega_Bh/\sqrt{\Omega_0}]^{-2/3}$.

\begin{figure}
\vspace{7.25in}
\caption[spectrum]{(a)  COBE FIRAS measurements of the CBR
temperature; (b) Summary of other CBR temperature measurements
(from \cite{FIRAS}); the dotted curve indicates the data
from the other high-precision measurement, by the UBC rocket-borne
COBRA instrument \cite{COBRA}.}
\end{figure}

The temperature of the CBR is very uniform across the sky, to better than
a part in $10^4$ on angular scales from tens of arcseconds
to 90 degrees; see Fig.~3.  Three forms of temperature anisotropy---two
spatial and one temporal---have now been detected:
(1) A dipole anisotropy of about a part in
$10^3$, generally believed to be due to the motion
of galaxy relative to the cosmic rest frame, at a speed
of about $620\km \sec^{-1}$ \cite{dipole}; (2) A yearly modulation in
the temperature in a given direction on the sky of
about a part in $10^4$, due to our orbital motion
around the sun at $30\km \sec^{-1}$, see Fig.~4 \cite{yearly}; and (3)
The temperature anisotropies detected by the Differential
Microwave Radiometer (DMR) on the Cosmic
Background Explorer (COBE) satellite,
$\langle (\Delta T /T)^2 \rangle_{10^\circ}^{1/2} =
1.1 \pm 0.2 \times 10^{-5}$ and $(\Delta T/T)_Q
= 6\pm 2 \times 10^{-6}$, where the first
measurement refers to the {\it rms} temperature fluctuation
averaged over the entire sky as measured by a beam of width $10^\circ$,
and the second is the magnitude of the
quadrupole temperature anisotropy \cite{DMR}.
The $10^\circ$ and quadrupole
anisotropies provide strong evidence for primeval
density inhomogeneities of the same
magnitude, which amplified by gravity,
grew into the structures that
we see today:  galaxies, clusters of galaxies,
superclusters, voids, walls, etc.

\begin{figure}
\vspace{3.5in}
\caption[anisotropy]{Summary of recent high-sensitivity
CBR anisotropy measurements; with the exception of
COBE all results are upper limits (from \cite{davisetal,wright}).
The solid boxes (MIT balloon experiment) have recently been reanalyzed
and shown to be a detection which is consistent with the COBE DMR
result \cite{meyeretal}.}
\end{figure}

\begin{figure}
\vspace{4in}
\caption[earth]{Yearly modulation of the CBR temperature---the
earth really orbits the sun(!) (from \cite{yearly}).}
\end{figure}

\subsubsection{Primordial nucleosynthesis}

Last, but certainly not least, there are the abundance
of the light elements.  According to the standard
cosmology, when the age of the Universe was measured
in seconds, the temperatures were of order MeV, and
the conditions were right for nuclear reactions
which ultimately led to the synthesis of significant amounts
of D, $^3$He, $^4$He, and $^7$Li.
The yields of primordial nucleosynthesis depend
upon the baryon density, quantified as the baryon-to-photon
ratio $\eta$, and the number of very light ($\la \MeV$)
particle species, often quantified as the equivalent number of
light neutrino species, $N_\nu$.  The predictions for
the primordial abundances of all four light elements
agree with their measured abundances provided that
$3\times 10^{-10} \la \eta \la 5\times 10^{-10}$ and
$N_\nu \la 3.4$; see Fig.~5 \cite{walkeretal}.

\begin{figure}
\vspace{7in}
\caption[bbn]{Predicted light-element abundances
and inferred abundances (from \cite{walkeretal}).
The measured primordial abundances are indicated
and the concordance region is shaded.}
\end{figure}

Accepting the success of the standard model of nucleosynthesis,
our precise knowledge of the
present temperature of the Universe allows us to
convert $\eta$ to a mass density, and by dividing
by the critical density, $\rho_{\rm crit} \simeq
1.88h^2 \times 10^{-29}\gcmm3$, to the fraction
of critical density contributed by ordinary matter:
\begin{equation}
0.011 \la \Omega_B h^2 \la 0.019;\qquad\Rightarrow
\ \   0.011 \la \Omega_B \la 0.12;
\end{equation}
this is the most accurate determination of the baryon
density.  Note, the uncertainty in the value of the
Hubble constant leads to most of the uncertainty in $\Omega_B$.

The nucleosynthesis bound to $N_\nu$, and more generally
to the number of light degrees of freedom in thermal
equilibrium at the epoch of nucleosynthesis, is consistent
with precision measurements of the properties of the $Z^0$
boson, which give $N_\nu = 3.0 \pm 0.05$; further, the
cosmological bound predates
these accelerator measurements!  The nucleosynthesis
bound provides a stringent limit to the existence
of new, light particles (even beyond neutrinos), and
even provides a bound to the mass the tau neutrino,
excluding a tau-neutrino mass between $0.5\MeV$ and
$25\MeV$ \cite{tau}.  Primordial nucleosynthesis provides
a beautiful illustration of the powers of the Heavenly
Laboratory, though it is outside the focus of these lectures.

The remarkable success of primordial nucleosynthesis gives
us confidence that the standard cosmology provides
an accurate accounting of the Universe at least as
early as $0.01\sec$ after the bang, when the temperature
was about $10\MeV$.

\subsubsection{Et cetera---and the age crisis?}

There are additional lines of reasoning and evidence
that support the standard cosmology \cite{dnsnature}.
I mention two:
the age of the Universe and structure formation.
I discuss the basics of structure formation a bit
later; for now it suffices to say that the standard
cosmology provides a basic framework for understanding
the formation of structure---gravitational instability---which
has recently been confirmed by COBE \cite{DMR}.
Here I will focus on the age of the Universe.

The expansion age of the Universe---time back to
zero size---depends upon the present expansion rate,
energy content, and equation of state:
$t_{\rm exp} = f(\rho ,p)H_0^{-1}\simeq
9.8h^{-1} f(\rho ,p)\Gyr$.  For a matter-dominated Universe,
$f$ is between 1 and 2/3 (for $\Omega_0$ between 0 and 1),
so that the expansion age is somewhere between $7\Gyr$ and
$20\Gyr$.  There are other independent measures of the
age of the Universe, e.g., based upon long-lived radioisotopes,
the oldest stars, and the cooling of white dwarfs.  These
``ages,'' ranging from 13 to 18 Gyr, span the
same interval(!).  This wasn't always
the case; as late as the early 1950's it was believe that
the Hubble constant was $500\km \sec^{-1} \Mpc^{-1}$, implying
an expansion age of at most $2\Gyr$---less than the age of the earth.
This discrepancy was an important motivation for the steady-state cosmology.

While there is {\it general} agreement between the expansion age
and other determinations of the age of the Universe, some
cosmologists are worried that cosmology is on the verge
of another age crisis \cite{distance2}.
Let me explain, while Sandage and
a few others continue to obtain values for the Hubble constant
around $50\kms \Mpc^{-1}$ \cite{h50}, a variety of different techniques
seem to be converging on a value around $80\pm 10\kms \Mpc^{-1}$
\cite{distance2}.
If $H_0 =80\kms\Mpc^{-1}$, then $t_{\rm exp} =
12f(\rho ,p) \Gyr$, and for $\Omega_0 =1$,
$t_{\rm exp} =8\Gyr$, which is clearly inconsistent
with other measures of the age.  {\it If} $H_0=80\kms\Mpc^{-1}$,
one is almost forced to consider the radical alternative of
a cosmological constant.  For example, even with
$\Omega_0=0.2$, $f\simeq 0.85$, corresponding to
$t_{\rm exp} \simeq 10\Gyr$; on the other hand,
for a flat Universe with $\Omega_\Lambda
=0.8$, $f\simeq 1.1$ and the expansion age $t_{\rm exp}
\simeq 13.5\Gyr$.  As I shall discuss later, structure formation
provides another motivation for a cosmological constant.
My own gut-level feeling is that when the dust settles,
we will find that $H_0=50\kms\Mpc^{-1}$; then again, maybe not.

\subsection {Basics of the Big Bang Model}

The standard cosmology is based upon the maximally spatially symmetric
Robertson-Walker line element
\begin{equation}
ds^2 = dt^2 -R(t)^2\left[ {dr^2\over 1-kr^2} +r^2
        (d\theta^2 + \sin^2\theta\,d\phi^2 ) \right];
\end{equation}
where $R(t)$ is the cosmic-scale factor, $R_{\rm curv}\equiv
R(t)|k|^{-1/2}$ is the curvature radius, and $k/|k| = -1,
0, 1$ is the curvature signature.  All three models are
without boundary:  the positively curved model is finite
and ``curves'' back on itself; the negatively curved
and flat models are infinite in extent (though finite
versions of both can be constructed by imposing a
periodic structure:  identifying all points in
space with a fundamental cube).  The Robertson-Walker
metric embodies the observed isotropy and homogeneity of
the Universe.  It is interesting to note
that this form of the line element was originally introduced
for sake of mathematical simplicity; we now know that
it is well justified at early times or today on large
scales ($\gg 10\Mpc$), at least within our Hubble volume.

The coordinates, $r$, $\theta$, and $\phi$, are referred
to as comoving coordinates:  A particle at rest in these
coordinates remains at rest, i.e., constant $r$, $\theta$,
and $\phi$.  A freely moving particle eventually comes
to rest these coordinates, as its momentum is red shifted
by the expansion, $p \propto R^{-1}$.
Motion with respect to the comoving coordinates (or cosmic
rest frame) is referred to as peculiar velocity; unless
``supported'' by the inhomogeneous distribution of matter
peculiar velocities decay away as $R^{-1}$.  Thus the
measurement of peculiar velocities, which is not easy
as it requires independent measures of both the distance
and velocity of an object, can be used to probe the
distribution of mass in the Universe.

Physical separations (i.e.,
measured by meter sticks) between freely moving particles
scale as $R(t)$; or said another way the physical separation
between two points is simply $R(t)$ times the coordinate
separation.  The momenta of freely propagating particles
decrease, or ``red shift,'' as $R(t)^{-1}$, and thus the
wavelength of a photon stretches as $R(t)$, which is
the origin of the cosmological red shift.  The red shift
suffered by a photon emitted from a distant galaxy
$1+z = R_0/R(t)$; that is, a galaxy whose light is
red shifted by $1+z$, emitted that light when the Universe
was a factor of $(1+z)^{-1}$ smaller.  Thus, when the
light from the most distant quasar yet seen ($z=4.9$) was emitted
the Universe was a factor of almost six smaller; when
CBR photons last scattered the Universe was about $1100$ times smaller.

\subsubsection{Friedmann equation and the First Law}

The evolution of the cosmic-scale factor is governed
by the Friedmann equation
\begin{equation}
H^2 \equiv \left({\dot R \over R}\right)^2 =
        {8\pi G \rho_{\rm tot} \over 3} - {k\over R^2};
\end{equation}
where $\rho_{\rm tot}$ is the total energy density of the
Universe, matter, radiation, vacuum energy, and so on.
A cosmological constant is often written as an additional
term ($=\Lambda /3$) on the rhs; I will choose to
treat it as a constant energy density (``vacuum-energy density''), where
$\rho_{\rm vac} = \Lambda /8\pi G$.  (My convention in this regard
is not universal.)  The evolution of the energy
density of the Universe is governed by
\begin{equation}
d(\rho R^3) = -pdR^3;
\end{equation}
which is the First Law of Thermodynamics for
a fluid in the expanding Universe.
(In the case that the stress energy of the Universe is comprised
of several, noninteracting components, this relation applies
to each separately; e.g., to the matter and radiation separately
today.)  For $p=\rho /3$, ultra-relativistic matter,
$\rho \propto R^{-4}$; for $p=0$, very nonrelativistic
matter, $\rho \propto R^{-3}$; and for $p=-\rho$, vacuum
energy, $\rho = \,$const.  If the rhs of the Friedmann
equation is dominated by a fluid
with equation of state $p = \gamma \rho$, it follows
that $\rho \propto R^{-3(1+\gamma )}$
and $R\propto t^{2/3(1+\gamma )}$.

We can use the Friedmann equation to relate the
curvature of the Universe to the energy density and
expansion rate:
\begin{equation}
{k /R^2 \over H^2} = \Omega -1;\qquad
\Omega = {\rho_{\rm tot}\over \rho_{\rm crit}};
\end{equation}
and the critical density today $\rho_{\rm crit}
= 3H^2 /8\pi G = 1.88h^2\gcmm3 \simeq 1.05\times 10^{4}
\eV \cmm3$.  There is a one to one correspondence
between $\Omega$ and the spatial curvature of the Universe:
positively curved, $\Omega_0 >1$; negatively curved, $\Omega_0
<1$; and flat ($\Omega_0 = 1$).  Further, the ``fate of the
Universe'' is determined by the curvature:  model universes
with $k\le 0$ expand forever, while those with $k>0$ necessarily
recollapse.  The curvature radius of the Universe is related
to the Hubble radius and $\Omega$ by
\begin{equation}
R_{\rm curv} = {H^{-1}\over |\Omega -1|^{1/2}}.
\end{equation}
In physical terms, the curvature radius sets the scale for
the size of spatial separations where
the effects of curved space become
``pronounced.''  And in the case of the positively curved
model it is just the radius of the 3-sphere.

The energy content of the Universe consists of matter
and radiation (today, photons and neutrinos).  Since
the photon temperature is accurately known,
$T_0=2.73\pm 0.01\,$K, the
fraction of critical density contributed by radiation
is also accurately known:  $\Omega_{\rm rad}h^2 = 4.18 \times
10^{-5}$.  The matter content is another matter.

\subsubsection {A short diversion concerning the present mass density}

The matter density today, i.e., the value of $\Omega_0$,
is not nearly so well
known \cite{dm}.  Stars contribute less than 1\% of critical density;
based upon nucleosynthesis, we can infer that baryons
contribute between 1\% and 10\% of critical.  The
dynamics of various systems allow astronomers to
infer their gravitational mass.  With their telescopes
they measure the amount of light, and form a mass-to-light
ratio.  Multiplying this by the measured luminosity
density of the Universe gives a determination of the
mass density.  (The critical mass-to-light ratio is
$1200h\,M_\odot /{\cal L}_\odot$.)

The motions of stars and gas clouds in spiral galaxies indicate that
most of the mass of spiral galaxies exists in
the form of dark (i.e., no detectable radiation),
extended halos, whose full extent
is still not known.  Many cite the
flat rotation curves of spiral galaxies,
which indicate that the halo density decreases
as $r^{-2}$, as the best evidence
that most of the matter in the Universe is dark.
Taking the mass-to-light ratio inferred for
spiral galaxies to be typical of the Universe as a
whole and remembering that the full extent of
the dark matter halos is not known,
one infers $\Omega_{\rm halo} \ga 0.03 - 0.1$.

The masses of clusters of galaxies can be estimated
using the virial theorem, and these mass estimates
too indicate the presence of large amounts of
dark matter.  Taking cluster mass-to-light ratios
to be typical of the Universe as a whole, in spite
of the fact that only about 1 in 10 galaxies resides
in a cluster, one infers $\Omega_{\rm cluster} \sim 0.1-0.3$.

Most galaxies are found in associations of a few galaxies
known as small groups.  Estimating the masses
of these systems using dynamics is tricky because
of the problem of ``interlopers,'' galaxies that happen
to be in the same part of the sky, but are not associated with
the group \cite{interlopers}.  This Fall, however, ROSAT detected the
weak x-ray emission from the hot gas in the small group
NGC 2300 \cite{smallgroup}; from their measurements
they were able to infer the shape of the gravitational
potential---and hence total mass of the group---as
well as the mass of the x-ray emitting gas and the visible mass in
galaxies.  They found that the total mass of the group
was about 20 times that in ordinary matter(!).  If one
takes this to be a universal ratio of the total amount
of matter to that in baryons and $\Omega_B\sim 0.05$,
one concludes that $\Omega_0 \sim 1$.

Not one of these methods is wholly satisfactory:
Rotation curves of spiral galaxies are still ``flat''
at the last measured points, indicating that the
mass is still increasing; likewise, cluster virial
mass estimates are insensitive to material that lies
beyond the region occupied by the visible galaxies---and
moreover, only about one galaxy in ten resides in
a cluster.   What one would like is a measurement of the mass of a
very big sample of the Universe, say a cube of
$100h^{-1}\Mpc$ on a side, which contains tens of thousands  of galaxies.

Over the past five years or so progress
has been made toward such a measurement.  It involves
the peculiar motion of our own galaxy, at a speed of about
$620\km\sec^{-1}$ in the general direction of
Hydra-Centaurus.  This motion is due to the lumpy
distribution of matter in our vicinity.  By using
gravitational-perturbation theory (actually, not much more
than Newtonian physics) and the distribution of galaxies in our vicinity
(as determined by the IRAS catalogue of infrared selected
galaxies), one can infer the average mass density in
a very large volume and thereby $\Omega_0$.

The basic physics behind the method is simple: the
net gravitational pull on our galaxy depends both upon
how inhomogeneous the distribution of galaxies is
and how much mass is associated with each galaxy;
by measuring the distribution of galaxies and our
peculiar velocity one can infer the ``mass per galaxy''
and $\Omega_0$.

The value that has been inferred is big(!):  $\Omega_{\rm IRAS}
\sim 1 \pm 0.2$ \cite{irasomega}.  Moreover,
the measured peculiar velocities of
other galaxies in this volume, more than thousand,
have been used in a similar manner and indicate
a similarly large value for $\Omega_0$ \cite{potent}.
While this technique is very powerful,
it does have its drawbacks:  One has to make simple
assumptions about how accurately
mass is traced by light (the observed galaxies);
one has to worry whether or not a significant portion
of our galaxy's velocity is due to galaxies outside the
IRAS sample---if so, this would lead to an overestimate
of $\Omega_0$; and so on.  This technique is not only
very promising---but provides the ``correct'' answer
(in my opinion!).

The so-called classical kinematic tests---Hubble diagram,
angle-red shift relation, galaxy count-red shift relation---can,
in principle, provide a determination of $\Omega_0$
\cite{Sandage}.  However, all these methods
require standard candles, rulers, or galaxies, and for this reason
have proved inconclusive.  However,
that hasn't stopped efforts to use these tests, particularly
the galaxy number-count test \cite{lohspillar}, and one or more of these
classical tests may one day provide a definitive measurement.

To summarize this aside on the mass density of the
Universe:
\begin{enumerate}

\item{}  Most of the matter is dark.

\item{}  Baryons provide between about 1\% and 10\% of the mass
density.

\item{}  $\Omega_0$ could conceivably be as small
as 0.1---in which case all the dark matter could be
baryons (e.g., neutron stars, ``jupiters,'' and so on).

\item{}  If asked for the value of $\Omega_0$,
a typical astronomer would respond with a
number in the interval $0.2\pm 0.1$.

\item{}  The evidence continues to mount
for a gap between $\Omega_B$ and $\Omega_0$---in which
case nonbaryonic dark matter is required.

\end{enumerate}

The current prejudice---and certainly that of this author---is
a flat Universe ($\Omega_0 = 1$) with nonbaryonic
dark matter, $\Omega_X\sim 1\gg \Omega_B$.  However, I shall continue
to display the $\Omega_0$ dependence of important quantities.

\subsubsection{The early, radiation-dominated Universe}

In any case, at present, matter outweighs radiation
by a wide margin.  However, since the energy density
in matter decreases as $R^{-3}$, and that
in radiation as $R^{-4}$ (the extra factor due
to the red shifting of the energy of relativistic
particles), at early times the
Universe was radiation dominated---indeed the calculations
of primordial nucleosynthesis provide excellent evidence for this.
Denoting the epoch of matter-radiation equality
by subscript `EQ,' and using $T_0=2.73\,$K, it follows that
\begin{equation}
R_{\rm EQ} = 4.18\times 10^{-5}\,(\Omega_0 h^2)^{-1};\qquad
T_{\rm EQ} = 5.62 (\Omega_0 h^2)\eV;
\end{equation}
\begin{equation}
t_{\rm EQ} = 4.17 \times 10^{10}(\Omega_0 h^2)^{-2}\sec .
\end{equation}
At early times the expansion rate and age of the Universe were
determined by the temperature of the Universe and
the number of relativistic degrees of freedom:
\begin{equation}
\rho_{\rm rad} = g_*(T){\pi^2T^4 \over 30}; \qquad
H\simeq 1.67g_*^{1/2} T^2 /\mpl ;
\end{equation}
\begin{equation}
\Rightarrow R\propto t^{1/2}; \qquad
t \simeq 2.42\times 10^{-6} g_*^{-1/2}(T/\GeV )^{-2}\,\sec ;
\end{equation}
where $g_* (T)$ counts the number of ultra-relativistic
degrees of freedom ($\approx$ the sum of the internal
degrees of freedom of particle species much less massive
than the temperature) and $\mpl \equiv G^{-1/2} = 1.22
\times 10^{19}\GeV$ is the Planck mass.  For example,
at the epoch of nucleosynthesis, $g_* = 10.75$ assuming
three, light ($\ll \MeV$) neutrino species; taking into
account all the species in the standard model,
$g_* = 106.75$ at temperatures much greater than $300\GeV$; see Fig.~6.

\begin{figure}
\vspace{4in}
\caption[g*]{The total effective number of relativistic
degrees of freedom $g_*(T)$ in the standard model of particle physics
as a function of temperature.}
\end{figure}

A quantity of importance related to $g_*$ is the
entropy density in relativistic particles,
$$s= {\rho +p \over T} = {2\pi^2\over 45}g_* T^3 ,$$
and the entropy per comoving volume,
$$S \ \  \propto\ \  R^3 s\ \  \propto\ \   g_*R^3T^3 .$$
By a wide margin most of the entropy
in the Universe exists in the radiation bath.
The entropy density is proportional
to the number density of relativistic particles.
At present, the relativistic particle species
are the photons and neutrinos, and the
entropy density is a factor
of 7.04 times the photon-number density:
$n_\gamma =413 \cmm3$ and $s=2905 \cmm3$.

In thermal equilibrium---which provides a good description
of most of the history of the Universe---the entropy per comoving
volume $S$ remains constant.  This fact is very useful.
First, it implies that the temperature and scale
factor are related by
\begin{equation}
T\propto g_*^{-1/3}R^{-1},
\end{equation}
which for $g_*=\,$const leads to the familiar $T\propto R^{-1}$.

Second, it provides a way of quantifying the net baryon
number (or any other particle number) per comoving volume:
\begin{equation}
N_B \equiv R^3n_B = {n_B\over s} \simeq (4-7)\times 10^{-11}.
\end{equation}
The baryon number of the Universe tells us two things:
(1) the entropy per particle in the Universe
is extremely high, about
$10^{10}$ or so compared to about $10^{-2}$ in the sun
and a few in the core of a newly formed neutron star.
(2)  The asymmetry between matter and antimatter is
very small, about $10^{-10}$, since at early times
quarks and antiquarks were roughly as abundant as
photons.  One of the great successes of particle
cosmology is baryogenesis, the idea that $B$, $C$, and
$CP$ violating interactions occurring out-of-equilibrium
early on allow the
Universe to develop a net baryon number of this magnitude
\cite{baryo}.

Finally, the constancy of the entropy per comoving
volume allows us to characterize the size of comoving
volume corresponding to our present Hubble volume in
a very physical way:  by the entropy it contains,
\begin{equation}
S_{U} = {4\pi\over 3}H_0^{-3}s \simeq 10^{90}.
\end{equation}

\subsubsection{The earliest history}

The standard cosmology is tested back to times as early
as about 0.01 sec; it is only natural to ask how far back
one can sensibly extrapolate.  Since the fundamental
particles of Nature are point-like quarks and leptons
whose interactions are perturbatively weak at energies much greater
than $1\GeV$, one can imagine extrapolating
as far back as the epoch where general relativity
becomes suspect, i.e., where quantum gravitational
effects are likely to be important:  the Planck epoch,
$t\sim 10^{-43}\sec$ and $T\sim 10^{19}\GeV$.
Of course, at present, our firm understanding
of the elementary particles and their interactions
only extends to energies of the order of $100\GeV$,
which corresponds to a time of the order of
$10^{-11}\sec$ or so.  We can be relatively certain
that at a temperature of $100\MeV -200\MeV$ ($t\sim 10^{-5}\sec$)
there was a transition (likely a second-order
phase transition) from quark/gluon
plasma to very hot hadronic matter, and that some
kind of phase transition associated
with the symmetry breakdown of the electroweak theory
took place at a temperature of the order of $300\GeV$
($t\sim 10^{-11}\sec$).

It is interesting to look at the progress that
has taken place since Weinberg's classic text on
cosmology was published in 1972 \cite{Weinberg}; at that time many
believed that the Universe had a limiting temperature
of the order of several hundred $\MeV$, due to the
exponentially rising number of particle states,
and that one could not speculate about earlier
times.  Today, based upon our present knowledge of physics
and powerful mathematical tools (e.g., gauge
theories, grand unified theories, and superstring
theory) we are able to make quantitative
speculations back to the Planck epoch---and
even earlier.  Of course, these speculations could be totally
wrong, based upon a false sense of confidence (arrogance?).
As I shall discuss, inflation is one of these
well defined---and well motivated---speculations
about the history of the Universe well after the
Planck epoch, but well before primordial nucleosynthesis.

\subsubsection{The matter and curvature dominated epochs}

After the equivalence epoch, the matter density exceeds
that of radiation.  During the matter-dominated epoch
the scale factor grows as $t^{2/3}$ and the age of the
Universe is related to red shift by
\begin{equation}
t = 2.06\times 10^{17} (\Omega_0 h^2)^{-1/2}(1+z)^{-3/2}\sec .
\end{equation}

If $\Omega_0 < 1$, the matter-dominated epoch is
followed by a ``curvature-dominated'' epoch where the
rhs of the Friedmann equation is dominated by the $|k|/R^2$ term.
When the Universe is curvature dominated it is said to
expand freely, no longer decelerating since the gravitational
effect of matter has become negligible:  $\ddot R\approx
0$ and $R\propto t$.
The epoch of curvature dominance
begins when the matter and curvature terms are equal:
\begin{equation}
R_{\rm CD} = {\Omega_0\over 1- \Omega_0} \longrightarrow \Omega_0;
        \qquad z_{\rm CD} = \Omega_0^{-1} - 2 \longrightarrow \Omega_0^{-1};
        \end{equation}
where the limits shown are for $\Omega_0\rightarrow 0$.
By way of comparison, in a flat Universe with a cosmological
constant, the Universe becomes ``vacuum dominated''
when $R=R_{\rm vac}$:
\begin{equation}
R_{\rm vac} = \left( {\Omega_0 \over 1-\Omega_0} \right)^{1/3}
\longrightarrow \Omega_0^{1/3}; \qquad z_{\rm vac}
= \left( {1-\Omega_0 \over \Omega_0}\right)^{1/3} -1
\longrightarrow \Omega_0^{-1/3} .
\end{equation}
For a given value of $\Omega_0$, the transition occurs
much more recently, which has important
implications for structure formation since small
density perturbations only grow during the matter-dominated era.

\subsubsection{One last thing:  horizons}

In spite of the fact that the Universe was
vanishingly small at early times, the rapid expansion
precluded causal contact from being established throughout.
Photons travel on null paths characterized
by $dr=dt/R(t)$; the physical distance that a photon
could have traveled since the bang until time $t$, the
distance to the horizon, is
\begin{eqnarray}
d_H(t) & = & R(t)\int_0^t {dt^\prime\over R(t^\prime )} \nonumber\\
       & = & t/(1-n) = nH^{-1}/(1-n) \qquad {\rm for}\ R(t)
        \propto t^n, \ \ n<1 .
\end{eqnarray}
Note, in the standard cosmology the distance to the
horizon is finite, and up to numerical factors,
equal to the age of the Universe or the Hubble radius, $H^{-1}$.
For this reason, I will use horizon and Hubble radius
interchangeably.\footnote{In inflationary models
the horizon and Hubble radius are not roughly equal
as the horizon distance grows exponentially relative
to the Hubble radius; in fact, at the end of inflation
they differ by $e^N$, where $N$ is the number of
e-folds of inflation.  However, I will slip and use
``horizon'' and ``Hubble radius'' interchangeably, though
I will always mean Hubble radius.}

An important quantity is
the entropy within a horizon volume:  $S_{\rm HOR}
\sim H^{-3}T^3$; during the radiation-dominated epoch
$H\sim T^2/\mpl$, so that
\begin{equation}
S_{\rm HOR} \sim \left( {\mpl\over T} \right)^3;
\end{equation}
from this we conclude that at early times the comoving
volume that encompasses all that we can see today
(characterized by an entropy of $10^{90}$) was comprised
of a very large number of causally disconnected regions.

\subsection {The challenge:  development of structure}

This brings us to what I believe is the major challenge
of the standard cosmology at present:  a detailed understanding
of the formation of structure in the Universe.
We have every indication that the Universe at early
times, say $t\ll 300,000\yrs$, was very homogeneous;
however, today inhomogeneity (or structure) is ubiquitous:
stars ($\delta\rho /\rho \sim 10^{30}$),
galaxies ($\delta\rho /\rho \sim 10^{5}$),
clusters of galaxies ($\delta\rho /\rho \sim 10-10^{3}$),
superclusters, or ``clusters of clusters''
($\delta\rho /\rho \sim 1$), voids ($\delta\rho /\rho
\sim -1$), great walls, and so on.

For some 25 years the standard cosmology has provided
a general framework for understanding this:
Once the Universe becomes matter dominated (around 1000 yrs
after the bang) primeval density
inhomogeneities ($\delta\rho /\rho \sim 10^{-5}$)
are amplified by gravity and
grow into the structure we see today \cite{sf}.
The fact that a fluid of
self-gravitating particles is unstable to the growth of
small inhomogeneities was first pointed out by Jeans
and is known as the Jeans instability.  The
existence of these inhomogeneities was confirmed in spectacular
fashion by the COBE DMR discovery of CBR anisotropy this
past spring:  The temperature
anisotropies detected almost certainly owe their
existence to primeval density inhomogeneities, as
causality precludes microphysical processes
from producing anisotropies on angular scales larger
than about $1^\circ$, the angular size of the horizon
at last scattering.

At last, the basic picture
has been put on firm ground (whew!).
Now the challenge is to fill in the details---origin
of the density perturbations, precise evolution of
the structure, and so on.  As I shall emphasize,
such an understanding may well be within reach,
and offers a window on the early Universe.

\subsubsection{The general picture:  gravitational instability}

Let us begin by expanding the
perturbation to the matter density in plane waves
\begin{equation}
{\delta\rho_M ({\bf x}, t)\over\rho_M} = {1\over (2\pi )^3}\int d^3k\,
	\delta_k(t) e^{-i{\bf k}
	\cdot {\bf x}},
\end{equation}
where $\lambda = 2\pi /k$ is the comoving wavelength
of the perturbation and $\lambda_{\rm phys} =
R\lambda$ is the physical wavelength.
The comoving wavelengths of perturbations corresponding
to bright galaxies, clusters, and the present horizon scale
are respectively:  about $1\Mpc$, $10\Mpc$, and $3000h^{-1}\Mpc$,
where $1\Mpc \simeq 3.09\times 10^{24}\rcm
\simeq 1.56\times 10^{38}\GeV^{-1}$.

The growth of small matter inhomogeneities of wavelength
smaller than the Hubble scale ($\lambda_{\rm phys} \la
H^{-1}$) is governed by a Newtonian equation:
\begin{equation}
{\ddot\delta}_k + 2H{\dot\delta}_k +v_s^2k^2\delta_k /R^2
= 4\pi G\rho_M \delta_k ,
\end{equation}
where $v_s^2 = dp /d\rho_M$ is the square of the sound
speed.  Competition between the pressure term and
the gravity term on the rhs determine whether or
not pressure can counteract gravity:
Perturbations with wavenumber larger than the Jeans wavenumber,
$k_J^2 = 4\pi GR^2 \rho_M /v_s^2$, are Jeans stable
and just oscillate; perturbations with smaller
wavenumber are Jeans unstable and can grow.
For cold dark matter $v_s \simeq 0$ and all scales are
Jeans unstable; even for baryonic matter, after decoupling
$k_J$ corresponds to a baryon mass of only about $10^5M_\odot$.
All the scales of interest here are Jeans unstable
and we will ignore the pressure term.

Let us discuss solutions to this equation under
different circumstances.  First, consider the Jeans
problem, evolution of perturbations in a static fluid,
i.e., $H=0$.  In this case Jeans unstable
perturbations grow exponentially,
$\delta_k \propto \exp (t/\tau )$ where $\tau = 1/\sqrt{4G\pi\rho_M}$.
Next, consider the growth of Jeans unstable perturbations in a
matter-dominated Universe,
i.e., $H^2=8\pi G\rho_M/3$ and $R\propto t^{2/3}$.  Because
the expansion tends to ``pull particles away from
one another,'' the growth is only power law,
$\delta_k \propto t^{2/3}$; i.e., at the same rate as the
scale factor.  Finally, consider a radiation or curvature
dominated Universe, i.e., $8\pi G\rho_{\rm rad}/3$ or
$|k|/R^2$ much greater than $8\pi G\rho_M/3$.  In this case, the expansion
is so rapid that matter perturbations grow very slowly,
as $\ln R$ in radiation-dominated epoch, or not at all
$\delta_k =\,$const in the curvature-dominated epoch.

The growth of nonlinear perturbations is another matter;
once a perturbation reaches an overdensity of order unity
or larger it ``separates'' from the expansion---i.e., becomes
its own self-gravitating system and ceases to expand
any further.  In the process of virial
relaxation, its size decreases by a factor of two---density
increases by a factor of 8; thereafter, its density contrast
grows as $R^3$ since the average
matter density is decreasing as $R^{-3}$, though smaller
scales could become Jeans unstable and collapse further to form
smaller objects of higher density, stars, etc.

{}From this we learn that structure formation begins when
the Universe becomes matter dominated and ends when
it becomes curvature dominated (at least the growth
of linear perturbations).  The total growth available for linear
perturbations is $R_{\rm CD}/R_{\rm EQ} \simeq 2.4\times 10^4\,
\Omega_0^2h^2$; since nonlinear structures have evolved
by the present epoch, we can infer that primeval perturbations of
the order $(\delta\rho_M/\rho_M)_{\rm EQ} \sim 4\times 10^{-5}\,
(\Omega_0 h)^{-2}$ are required.  Note that in a low-density
Universe larger initial perturbations are necessary as there
is less time for growth (``the low
$\Omega_0$ squeeze'').  Further, in a baryon-dominated Universe
things are even more difficult as perturbations in the baryons
cannot begin to grow until after decoupling since matter is tightly
coupled to the radiation.  (In a flat, low-$\Omega_0$ model
with a cosmological constant the growth of linear
fluctuations continues until almost today since $z_\Lambda
\sim \Omega_0^{-1/3}$, and so the total growth factor
is about $2.4\times 10^4 (\Omega_0h^2)$.  We will return to
this model later.)

\subsubsection{CBR temperature fluctuations}

The existence of density inhomogeneities has another important
consequence:  fluctuations in the temperature of the CBR of
a similar amplitude \cite{dt/t}.  The temperature difference measured
between two points separated by a large angle ($\ga 1^\circ$)
arises due to a very simple physical effect:\footnote{Large
angles mean those larger than the angle
subtended by the horizon-scale at decoupling, $\theta
\sim H_{\rm DEC}^{-1}/H_0^{-1}\sim z_{\rm DEC}^{-1/2}\sim 1^\circ$.}
The difference in the gravitational potential between the two points on
the last-scattering surface, which in turn is related
to the density perturbation, determines
the temperature anisotropy on the angular scale subtended
by that length scale,
\begin{equation}
\left({\delta T \over T}\right)_\theta =
-\left({\delta\phi \over 3}\right)_\lambda \approx {1\over 2}
\left( {\delta\rho\over \rho}\right)_{{\rm HOR},\lambda};
\end{equation}
where the scale $\lambda \sim 100h^{-1}\Mpc(\theta /{\rm deg})$
subtends an angle $\theta$ on the last-scattering
surface.  This is known as the Sachs-Wolfe effect \cite{SW}.

The quantity $(\delta\rho /\rho)_{{\rm HOR},\lambda}$ is the
amplitude with which a density perturbation crosses
inside the horizon, i.e., when $R\lambda \sim H^{-1}$.  Since
the fluctuation in the gravitational potential
$\delta \phi \sim (R\lambda /H^{-1})^2(\delta\rho /\rho )$,
the horizon-crossing amplitude is equal to the
gravitational potential (or curvature) fluctuation.
The horizon-crossing amplitude $(\delta\rho /\rho )_{\rm HOR}$
has several nice features:  (i) during the matter-dominated era
the potential fluctuation on a given scale remains constant, and
thus the potential fluctuations at decoupling on scales
that crossed inside the horizon after matter-radiation
equality, corresponding to angular scales $\la 0.1^\circ$, are just
given by their horizon-crossing amplitude;
(ii) because of its relationship to $\delta\phi$
it provides a dimensionless,
geometrical measure of the size of the density perturbation
on a given scale, and its effect on the CBR; (iii) by specifying
perturbation amplitudes at horizon crossing one can effectively
avoid discussing the evolution of density perturbations
on scales larger than the horizon, where a Newtonian analysis
does not suffice and where gauge subtleties (associated
with general relativity) come into play; and finally (iv) the
density perturbations generated in inflationary models
are characterized by $(\delta\rho /\rho )_{\rm HOR}\simeq
\,$const.

On angular scales smaller than about $1^\circ$ two other
physical effects lead to CBR temperature fluctuations:
the motion of the last-scattering surface (Doppler) and
the intrinsic fluctuations in the local photon temperature.
These fluctuations are much more difficult to compute, and
depend on microphysics---the ionization history of the Universe
and the damping of perturbations in the photon-baryon fluid
due to photon streaming.  Not only are the Sachs-Wolfe fluctuations
simpler to compute, but they accurately mirror the
primeval fluctuations since at the epoch of decoupling
microphysics is restricted to angular scales less than about a degree.

In sum, on large angular scales the
Sachs-Wolfe effect dominates; on the scale of about
$1^\circ$ the total CBR fluctuation is about twice that due
to the Sachs-Wolfe effect; on smaller scales the Doppler and
intrinsic fluctuations dominate.  CBR temperature fluctuations
on scales smaller than about $0.1^\circ$ are severely
reduced by the smearing effect of the finite thickness of
last-scattering surface.

Details aside, in the context of the gravitational instability
scenario density perturbations of sufficient amplitude
to explain the observed structure lead to temperature
fluctuations in the CBR of characteristic size,
\begin{equation}
{\delta T\over T} \approx 10^{-5}\,(\Omega_0 h)^{-2}.
\end{equation}
To be sure I have brushed over important details, but
this equation conveys a great deal.  First, the overall
amplitude is set by the inverse of the growth factor,
which is just the ratio of the radiation energy density
to matter density at present.  Next, it explains why
theoretical cosmologists were so relieved when
the COBE DMR detected temperature fluctuations of this amplitude,
and conversely why one heard offhanded remarks before
the COBE DMR detection that the standard cosmology
was in trouble because the CBR temperature was too uniform
to allow for the observed structure to develop.
Finally, it illustrates one of the reasons
why cosmologists who study structure formation
have embraced the flat-Universe model with such
enthusiasm:  If we accept the Universe that meets the eye, $\Omega_0\sim
0.1$ and baryons only, then the simplest models of structure
formation predict temperature fluctuations of the order
of $10^{-3}$, far too large to be consistent with observation.
Later, I will mention Peebles' what-you-see-is-what-you-get
model \cite{pib}, also known as PIB for primeval baryon isocurvature
fluctuation, which is still viable because the spectrum of perturbations
decreases rapidly with scale so that the perturbations that
give rise to CBR fluctuations are small (which
is no mean feat).  Historically, it was
fortunate that one started with a low-$\Omega_0$, baryon-dominated
Universe:  the theoretical predictions for the CBR fluctuations
were sufficiently favorable that experimentalist
were stirred to try to measure them---and then, slowly,
theorists lowered their predictions.  Had the theoretical
expectations begun at $10^{-5}$, experimentalists might
have been too discouraged to even try!

\subsubsection{An initial data problem}

With the COBE DMR detection in hand we can praise
the success of the gravitational instability
scenario; however, the details now remain to be
filled in.  The structure formation problem is now one
of initial data, namely
\begin{enumerate}
\item The quantity and composition of
matter in the Universe, $\Omega_0$, $\Omega_B$, and $\Omega_{\rm other}$.

\item The spectrum of initial density perturbations:
for the purist, $(\delta\rho /\rho )_{\rm EQ}$, or
for the simulator, the Fourier amplitudes
at the epoch of matter-radiation equality.

\end{enumerate}
In a statistical sense, these initial data
provide the ``blueprint'' for the formation of structure.

The initial data are the challenge and the opportunity.  Although
the gravitational instability picture has been around since
the discovery of the CBR itself, the lack of specificity
in initial data has impeded progress.  With the advent
of the serious study of the earliest history of the Universe
a new door was opened.  We now have several well motivated
early-Universe blueprints:  Inflation-produced density
perturbations and nonbaryonic dark matter; cosmic-string
produced perturbations and nonbaryonic dark matter \cite{cs}; texture
produced density perturbations and nonbaryonic dark matter \cite{texture};
a baryon-dominated Universe with isocurvature
fluctuations\footnote{Isocurvature baryon-number fluctuations correspond
at early times to fluctuations in the local baryon number
but not the energy density.  At late times, when the Universe
is matter dominated, they become fluctuations in the
mass density of a comparable amplitude.} \cite{pib}.
Structure formation also provides the
opportunity to probe the earliest history of the Universe,
by testing these interesting, if not bold, blueprints.
I will be focusing on the blueprints motivated by inflation.

\section {Inflation:  An Overview}

\subsection{Shortcomings of the Standard Cosmology}

By now the shortcomings of the standard cosmology are
well appreciated:  the horizon or large-scale smoothness problem;
the small-scale inhomogeneity problem (origin of density perturbations);
the flatness or oldness problem; and the monopole problem.
I will only briefly review them here.  They do not indicate any
logical inconsistencies of the standard cosmology; rather,
that very special initial data seem to be required for evolution
to a universe that is qualitatively similar to ours today.
Nor is inflation the first attempt to address these shortcomings:
Over the past two decades cosmologists have pondered this
question and proposed other solutions \cite{other}.
Inflation is a solution based upon well-defined,
albeit speculative, early-Universe microphysics describing
the post-Planck epoch.

The uniformity of the CBR temperature, to better than
a part in $10^4$, implies that the
Universe on the largest scales (say $\ga 100h^{-1}\Mpc$)
is very smooth as density inhomogeneities induce temperature
fluctuations of a similar magnitude.  The existence of particle
horizons in the standard cosmology precludes explaining
the smoothness as a result of microphysical events:  The
horizon at decoupling, the last time one could imagine
temperature fluctuations being smoothed by particle interactions,
corresponds to an angular scale on the sky of about
$1^\circ$, which precludes temperature variations on larger scales
from being erased.  In terms of entropy, the presently
observed Universe, corresponds to a comoving volume containing
an entropy of order $10^{90}$; during the early radiation
dominated epoch the horizon volume contained an entropy of
order $(\mpl /T)^3$, implying that at early times our current
Hubble volume consisted of countless, causally distinct regions.

To account for the small-scale lumpiness of the
Universe today, density perturbations with horizon-crossing
amplitudes of $10^{-5}$ on scales of $1\Mpc$ to $10^4\Mpc$ or so
are required.  As can be seen in Fig.~7, in the
standard cosmology the physical
size of a perturbation, which grows as the scale factor,
begins larger than the horizon and relatively late
in the history of the Universe crosses inside the horizon,
\begin{eqnarray}
t_{\rm HOR} & \simeq & 3\times 10^8\,(\lambda /\Mpc )^2\,\sec
        \qquad \lambda \la 13h^{-2}\Mpc ; \nonumber \\
            & \simeq & 3\times 10^7\,(\lambda /\Mpc )^3\,\sec
            \qquad  \lambda \ga 13h^{-2}\Mpc .
\end{eqnarray}
This precludes a causal microphysical explanation for
the origin of the required density perturbations.\footnote{Of
course, it is possible to produce the perturbations
at very late times, when the relevant scale has already
crossed inside the horizon \cite{latetime}; the motivation
for the nonstandard microphysics required to do so is
lacking at present.  It is also possible for microphysics to
produce isocurvature perturbations by producing a pressure
wave that eventually propagates to large scales;
this is the type of perturbation that is generated
by cosmic strings or textures.}

\begin{figure}
\vspace{4.5in}
\caption[size]{The physical wavelength
of a density perturbation and the horizon size $H^{-1}$
as a function of scale factor; $\lambda_{\rm GAL}$ indicates
a galactic sized perturbation ($\lambda\sim 1\Mpc$)
and $\lambda_{\rm HOR}$ corresponds to the present
Hubble radius (horizon).  Microphysics operates on scales
$\la H^{-1}$; without inflation scales cross the Hubble
radius but once.}
\end{figure}

The fact that $\Omega_0$ is of order unity means
that the curvature radius is comparable to the Hubble
radius.  Had that been the case at the initial epoch,
the Universe would be a very different place today:
Since the curvature term in the Friedmann equation
decreases only as $R^{-2}$, while the matter and radiation
densities decrease as $R^{-3}$ and $R^{-4}$ respectively,
a curvature radius comparable to the Hubble radius
early on would have led to a Universe that quickly
became curvature dominated.  For positive curvature,
recollapse would follow quickly, and for negative
curvature, a coasting phase that would lead to a
Universe that cools too quickly (for $t_{\rm initial}
\sim 10^{-43}\sec$, the temperature reaches $3\,$K
at an age of $10^{-11}\sec$).  Put another way,
$\Omega$ is an unstable fixed point:
\begin{eqnarray}
\Omega (t) & = & {1\over 1-x(t)} \nonumber \\
x(t) & = & {k/R^2\over 8\pi G\rho /3} ;
\end{eqnarray}
the deviation of $\Omega (t)$ from unity increases
as $x(t) \propto R^n$, $n=2$ (radiation-dominated epoch),
$n=1$ (matter-dominated epoch).  In order that
$\Omega$ still be close to unity today, it must have
extremely close to unity early on; for $t_{\rm initial}
\sim 10^{-43}\sec$, $|\Omega (t_{\rm initial}) -1|\la
10^{-60}$ is necessary.  Thus, for most of
its history the Universe must have been extremely flat,
i.e., $R_{\rm curv}\gg H^{-1}$; if $\Omega_0$ is not
equal to unity, then the Universe just today is
beginning to exhibit its curvature.  Why now?

Last, I mention the monopole problem:  The simplest grand
unified theories and the standard cosmological lead to
a disastrous prediction, the extreme overproduction of
magnetic monopoles \cite{monopole}.  This overproduction
traces to the smallness of the horizon at very early times:
magnetic monopoles are produced as defects of the GUT phase
transition at an abundance of about 1 per horizon volume
which corresponds to a present monopole to photon ratio
of order $(T_{\rm GUT}/\mpl )^3$.

The first three problems do not involve logical inconsistencies:
The initial data for a perturbed FRW model that is extremely
flat exist.  Rather, it is the fact that such initial data are
``very special'' which is disturbing.
Collins and Hawking quantified it:  The set of initial data
that evolve to a state qualitatively similar to our Universe
is of measure zero \cite{collins}.  Maybe the Creator had a lucky day!
Or better yet, perhaps the present state of the Universe traces to events
that took place early on.  Inflation provides an interesting
example of the latter.

\subsection{Generic Aspects of Inflation}

Inflationary cosmology has become a very mature
subject in the decade since Guth wrote his influential
paper \cite{guth} that launched the inflationary cosmology boom.
While there are a multitude of different
kinds of inflation (see below),
two features are common to all models of inflation \cite{hu}

\begin{itemize}

\item Superluminal expansion

\item Massive entropy production

\end{itemize}

Superluminal expansion refers to accelerated growth
of the scale factor ($\ddot R >0$ which implies
$R\propto t^n$ with $n>1$), and its necessity is
easy to understand.  In order
that the physical size of a comoving scale, $d_{\rm phys}
\propto R(t)$, begin sub-Hubble size and
and become super-Hubble size, $R(t)$ must increase
faster than $t$ since $H^{-1}\propto R(t)^{1/n}$.  Thus, ``superluminal''
expansion is a necessary kinematic requirement if one is to
both solve the horizon and create density perturbations
(see Fig.~7).

The reason for the second requirement is equally simple:
In the absence of entropy production the entropy per
comoving volume $S\propto (RT)^3$ remains constant;
rapid expansion can create a ``very large''
smooth patch, but the entropy
within that patch remains constant.  As discussed above,
at early times the entropy within a horizon-sized patch
is very small, too small to account for the entropy
within our present Hubble volume.  Only massive entropy
production can change this \cite{hu}.

To illustrate, consider Guth's original model
of inflation based upon a first-order phase transition \cite{guth}.
The basic idea is that the Higgs field responsible for
the spontaneous breakdown of the GUT symmetry gets
``hung up'' in a local, high-energy, minimum of its
potential (more precisely, free-energy density).
At high temperatures the state of minimum free
energy is characterized by $\phi = 0$, indicating
that the full GUT symmetry is manifest; as the temperature drops
below the critical temperature, the state of minimum
free energy is characterized by $\phi \not= 0$, the state
that exhibits broken symmetry.  In
a first-order transition $\phi = 0$ can remain
a local minimum of the free energy, separated from
the global minimum by a potential barrier; see Fig.~8.
During the time that $\phi$ is hung up
the large vacuum-energy density,
$\rho = V (\phi = 0)\equiv {\cal M}^4$, drives
very rapid expansion ($\cal M$ is
the energy scale that characterizes of the symmetry breaking).

\begin{figure}
\vspace{2.25in}
\caption[V_T]{The free-energy density as a function
of temperature for a first-order phase transition.}
\end{figure}

For definiteness, take ${\cal M} = 10^{14}\GeV$,
a typical scale for inflation; the Hubble time
associated with the false-vacuum energy $H^{-1} \sim
10^{-34}\sec$.  The size of a region that one might
expect to be smooth is of order $ct\sim 10^{-23}\rcm$;
the entropy within such a patch is
of order $10^{14}$.  While the Higgs field is trapped
in the false vacuum, the temperature of the Universe
continues to decrease as $R^{-1}$; very soon the thermal
energy density becomes insignificant compared to the
constant false-vacuum energy density.  At this point,
the Universe enters a de Sitter phase of exponential
expansion since $\rho \sim {\cal M}^4=\,$const;
this is the superluminal expansion.
As the Universe expands, it cools exponentially
with the entropy per comoving volume remaining constant;
the smooth horizon-sized patch continues to contain
an entropy of only $10^{14}$ as it grows exponentially in size.

During inflation the scale factor undergoes many e-folds;
the precise number is determined by how long the Higgs field
is hung up:  $N=H\Delta t$.  Again, for definiteness,
suppose that the Universe gets hung up for a mere
$10^{-32}\sec$; then, during inflation the patch
grows in linear size by $e^{100}\sim 10^{43}$ and
its temperature drops by the same factor.  Thus far,
inflation has done little.  When the Higgs
field does make its way to the true vacuum, the
enormous false-vacuum energy is released and ultimately
thermalized, reheating the patch to a temperature
of order ${\cal M}\sim 10^{14}\GeV$, thereby increasing
the entropy of the patch by a factor of $e^{3N}\sim
10^{129}$.  This is the massive entropy production.
After ``reheating'' the
patch contains an entropy of order $10^{143}$, and
can easily contain the comoving volume that corresponds
to our present Hubble radius, which is characterized
by an entropy of ``only'' $10^{90}$.

It is clear that the smoothness problem has been
solved.  The kinematic requirement for producing
density perturbations on astrophysically interesting
scales has been satisfied; the mechanism that produces
density perturbations, quantum fluctuations in the
$\phi$ field, will be discussed later.
What about the flatness
problem?  Suppose for definiteness that the curvature
radius at the beginning of inflation is of order
the Hubble radius (which corresponds to $\Omega$
just beginning to deviate from unity); at the
end of inflation the curvature radius has grown
by a factor of $e^N$, while the energy density
has remained constant.  This means that $\Omega_{\rm end}
= 1/[1-(k/R^2)/(8\pi G\rho /3)] \simeq
1 \pm e^{-2N}$ has been reset to a value exponentially close to
unity.  Using our fiducial numbers, at the end
of inflation the curvature radius is order
$10^{20}\rcm$; from then until today it grows
by a factor of ${\cal M}/3\,{\rm K} \sim 10^{27}$,
reaching a present size of order $10^{47}\rcm$.
This is enormous compared to the present Hubble
radius and implies that $\Omega$ is still very
close to unity today.  The flatness problem has clearly been solved
and a flat Universe predicted.

Consider the fate of monopoles---or any other
``cosmic pollutant'' in the pre-inflationary Universe.
The number of monopoles within the patch ($=N_M$) remains
constant; however, the number per comoving
volume, $n_M/s = N_M/S$, decreases by a factor of $e^{3N} \sim 10^{129}$
due to the massive entropy production.  Undesirables are
diluted away!  Of course, this also implies that
the baryon number of the Universe, $n_B/s \sim 10^{-10}$,
must be produced after inflation.

Finally, a simple exercise; what is the minimum amount
of inflation needed to solve the smoothness problem?
Start with a Hubble-sized patch at the beginning
of inflation; it contains an entropy of $S_{\rm initial}
\sim H^{-3} T^3 \sim (\mpl /{\cal M})^3$.  Assuming
perfect conversion of vacuum energy to radiation,
after inflation the entropy contained within the
patch is $e^{3N}S_{\rm initial} \sim e^{3N}\mpl^3/{\cal M}^3$.
To solve the smoothness problems this must be greater
than $10^{90}$, which implies
\begin{equation}
N\ga N_{\rm min} = 56 + \ln ({\cal M}/10^{14}\GeV ).
\end{equation}
Equivalently, one can express the size of the patch
today relative to the present Hubble radius,
\begin{equation}
d_{\rm patch} = \exp (N-N_{\rm min})\,H_0^{-1}.
\end{equation}

What about the flatness problem?  It is simple to show the
present value of $\Omega$ is related to that at the
beginning of inflation and the size of the patch today:
\begin{equation}
{|\Omega_0-1|} = \left( {H_0^{-1}\over d_{\rm patch}} \right)^2
        \,|\Omega_{\rm preinflation}-1|.
\end{equation}
Remarkably enough, the amount of inflation required to
solve the flatness and smoothness problems is the same.
Put another way, if one comfortably solves
the smoothness problem, $\Omega_0$ is necessarily very, very
close to unity.  This means that a flat
Universe is an unequivocal prediction of inflation.

\subsection{Current Status of Inflationary Models}
\subsubsection{Types of inflation}

In this very brief overview I
divide models of inflation into three broad classes:  old, slow rollover,
and first-order (or extended).  By old inflation I mean Guth's original model,
which I forgot to mention was a nonstarter!
Let me explain; once trapped in the false vacuum,
the Higgs field must quantum-mechanically tunnel to the
true vacuum; in order to ensure a sufficient
amount of inflation, this transition must not
occur until 60 or so Hubble times after inflation
has begun.  As we shall see this is essentially impossible to arrange.

The decay of the false vacuum is well
understood \cite{coleman}:  It proceeds via the nucleation
of bubbles of true vacuum that expand outward at the
speed of light.  For a given potential the bubble nucleation
rate (per unit volume) $\Gamma$ is straightforward to calculate
\cite{coleman}.  Roughly speaking, bubbles convert all of space
into the true vacuum when $\Gamma /H^4$, the
number of bubbles nucleated in a Hubble volume in a Hubble
time exceeds order unity; since each bubble nucleated
during a Hubble time liberates about a Hubble volume,
$\Gamma /H^4\sim 1$ ensures that all of space is converted
to true-vacuum in a Hubble time (before the expansion ``creates''
more false vacuum).  The false-vacuum energy is converted into ``heat''
by the collision of vacuum bubbles \cite{bubblerh}.

The recipe for successful old inflation
is for $\Gamma /H^4$ to remain less than
unity for 60 or so Hubble times and then increase to
greater than unity.  Unfortunately, shortly after inflation
begins $\Gamma$, like the expansion rate becomes
constant, as the temperature of the Universe rapidly approaches
zero and become irrelevant.  This is the
fundamental problem with old inflation; $\Gamma /H^4$ is
constant.  The Universe can either
get hung up in the false vacuum and inflate, or make the transition
to the true vacuum, not both!

Slow-rollover inflation solved this problem, but at a price.
The fix, suggested independently by Linde \cite{linde1},
and Albrecht and Steinhardt \cite{as}, is for inflation to occur as the
scalar field slowly rolls the potential.
They proposed using very flat potentials with small or nonexistent
barriers between the false and true vacuum states; the
vacuum-driven expansion takes place as the scalar field slowly
(timescale $\ga 60H^{-1}$), but inevitably rolls toward
the true-vacuum state.  When the scalar field responsible
for inflation (often called the inflaton) reaches the true
minimum of its potential it oscillates about it, the large
vacuum energy having been converted into coherent inflaton
oscillations.  These oscillations ultimately
decay into light-particle states reheating the Universe.
{}From the quantum view, these coherent field oscillations correspond
to zero-momentum inflaton particles; the decay of the scalar-field
oscillations corresponds to the decay of massive inflaton particles
\cite{reheat}.

Slow rollover led to the first viable models
of inflation.  There was, however, a price:  In all models
of slow-rollover the inflaton field must be very weakly
coupled (dimensionless self coupling of order $10^{-14}$
or so); as we shall see this is dictated by achieving
density perturbations of size $10^{-5}$ or so.
Because of this fact, the inflaton cannot be directly
responsible for GUT symmetry breaking as loop corrections
from the inflaton-gauge interaction would spoil the flatness
of the potential.  The decoupled nature of the scalar
field responsible for inflation gave birth to its name.
In the broadest sense, slow-rollover inflation refers
to any model of inflation where the a scalar field is
displaced from the minimum of its potential and slowly
rolls to the minimum.  The minimum can be away from
the origin, as with a potential associated
with spontaneous symmetry breaking (often referred
to as ``new inflation''), or at the origin,
e.g., $V(\phi )=\lambda\phi^4$ or $V(\phi )= m^2\phi^2/2$
(often referred to as chaotic inflation).

The latest and perhaps most interesting development in
inflationary models is first-order (or extended)
inflation \cite{foi}.  In many ways it combines the best features
of old inflation---intimate connection to particle physics
phenomenology---and slow-rollover inflation---it works!
As the name suggests, these models are associated with
a first-order phase transition; how then do these models
solve the Guth dilemma---the constancy of
$\Gamma /H^4$?  The first model of this
type was due to La and Steinhardt \cite{extended};
their new twist was to use the Brans-Dicke theory of
gravity rather than general relativity.  In Brans-Dicke the
gravitational constant $G_{\rm eff}
=\Phi^{-2}$, and evolves as the Brans-Dicke field $\Phi$
evolves.  Because of this, for constant energy density
{\it the scale factor only increases as a power of time,}
$R(t)\propto t^{\omega +1/2}$ {\it and $H$ decreases
with time;}  here $\omega$ is the coefficient of the
kinetic-energy term for
$\Phi$.  Thus, the efficiency of bubble nucleation
$\Gamma /H^4 \propto t^4$ increases during inflation;
at early times it can be much less than unity
(so that the Universe remains trapped in the false
vacuum) and then exceeds unity triggering the end
of inflation via the nucleation and percolation
of bubbles of true vacuum.

Models based on variations of this idea
have been proposed.  For example, if the Higgs field
couples to other fields which are evolving during inflation,
then $\Gamma$ will vary during inflation, leading to the variation
of $\Gamma /H^4$ \cite{twofield}.  In first-order inflation models
the Higgs field plays a relatively passive role,
remaining trapped in the false vacuum during inflation;
further, it need not be weakly coupled, nor is the shape
of its potential particularly relevant.

By means of a conformal transformation extended inflation
can be recast as slow-rollover inflation with an exponential
potential with $\ln \Phi$ field playing the role of the inflaton \cite{kst}.
In first-order inflation models there is another problem
one has to worry about:  If $\Gamma /H^4$ does not change
rapidly enough, then
too many bubbles will be nucleated long before the end
of inflation; these bubbles eventually
grow to astrophysical size and can have
disastrous consequences (large anisotropies in the CBR,
interference with primordial nucleosynthesis, and so on)
\cite{bigbubbles}.   To avoid ``the big-bubble problem''
in extended inflation $\omega$ must be less than
about 20; that there be an upper limit to $\omega$ is
not surprising since in the limit $\omega\rightarrow\infty$,
Brans-Dicke goes to general relativity.

\subsubsection{Viable models}

There is no standard model of inflation; nor is there a model of
inflation without some flaw.   There are
a number of ``proof of existence'' models, models
that successful implement inflation, but are only
beautiful in the eyes of their creators.  Of course,
this situation should be viewed in light of our general
ignorance about physics at energy scales $\gg10^3\GeV$
(most inflation models involve an energy scale of
order $10^{14}\GeV$).  Moreover, the same criticism---lack
of a standard model---applies to baryogenesis, and
applied to primordial nucleosynthesis until the early 1970's!

\bigskip
\noindent{\bf Slow-rollover.}
There are numerous viable models; I will mention
but a representative few.  There is an almost decade
old model based upon an ordinary GUT due to Shafi and Vilenkin \cite{sv},
and Pi \cite{Pi}.  This model has the virtue that the inflaton
field does more than cause inflation; it also breaks Peccei-Quinn
symmetry and induces GUT symmetry breaking (by producing a
negative mass-squared for the GUT Higgs field).  After inflation
the Universe reheats to a temperature of order $10^7\GeV$, and
a scenario for baryogenesis is included.  In short, it is a complete model.

In passing, let me mention a similar model just
proposed by Knox and myself \cite{ewinflation}.
The new twist is that the scale of inflation can be as
small as the electroweak scale(!), and the inflaton field
can be used to induce electroweak-symmetry breaking and
other low-energy phenomena (e.g., righthanded neutrino masses).
In principle, this model can be tested in laboratory experiments.
Of course, this model is only viable provided one believes
that the baryon asymmetry of the Universe can be
produced at the weak scale or below.

There are many supersymmetric implementations of slow-rollover
inflation \cite{olive}; a particularly elegant one is that of
Holman, Ramond, and Ross \cite{hrr}.  The superpotential
for their inflaton is very simple, $W(\phi ) = (\Delta^2/M)
(\phi -M)^2$; here $M=\mpl/\sqrt{8\pi}$ and
$\Delta$ is the GUT scale.  In this model, the
self coupling of the inflaton in its scalar
potential is given by the
fourth power of the ratio of the GUT to Planck scales,
$(\Delta /M)^4$, and the canonical small number
arises because of the discrepancy between the GUT
and Planck scales.
The reheat temperature in this model is order $10^6\GeV$,
and the details of baryogenesis are spelled out.

There is a model called (by the authors) ``natural inflation''
\cite{natural}.  The primary purpose of this model is
to address the small self-coupling of the inflaton.
To wit, the inflaton is a pseudo Nambu-Goldstone
boson akin to the axion; a Nambu-Goldstone boson has an
absolutely flat potential, i.e. is massless, and becomes
a pseudo Nambu-Goldstone boson due to explicit symmetry
breaking.  The potential,
$V(\phi ) = \Lambda^4[1 + \cos (\phi /f)]$, has two
energy scales:  $f \sim \mpl$, the scale of the spontaneous
symmetry breaking and $\Lambda \sim 10^{-5} f$, the scale of explicit
symmetry breaking (GUT scale?).
(In the axion analogy, $\Lambda = \Lambda_{\rm QCD}
\simeq 200\MeV$ and $f$ is the PQ symmetry-breaking scale.)
Some superstring adherents have
taken interest in this model as superstring theories
often have pseudo Nambu-Goldstone bosons with Planck-scale
symmetry breaking.

There is a broad class of slow-rollover models referred to as
chaotic inflation; they illustrate the simplicity
of inflation and were pioneered by Linde \cite{chaotic}.
The potentials for these models are not of the symmetry
breaking variety as the minima are at $\phi =0$; e.g.,
$V(\phi )= \lambda\phi^4$ or $V(\phi )=m^2\phi^2/2$.
In chaotic inflation, the inflaton field begins far from
the origin, $\phi \ga 5\mpl$---the further the better.
As with all slow-rollover models, there is a small, dimensionless
number:  $\lambda \sim 10^{-14}$ or $m^2\simeq 10^{-12}\mpl^2$.
No attempt is made to connect these models with particle physics.

There are models where the inflaton
field is not actually a scalar field; e.g., where
it is related to the size of the compactified dimensions
in models with extra dimensions \cite{witterich},
or is related to the scalar curvature $\cal R$ in higher
derivative theories of gravity) \cite{R2}.

The common undesirable feature of all slow-rollover models
is a small, dimensionless number of order $10^{-14}$,
typically the self coupling of the inflaton; as we shall
discuss, this small number is necessary to guarantee
density perturbations of the appropriate size.
To ensure the stability of the flatness of the potential against
quantum (radiative) corrections the inflaton must be
weakly coupled to the ``rest of the world,''  and in this
since, {\it all} the models mentioned are natural.  However,
weak coupling works at cross purposes with reheating and
baryogenesis.  Slow-rollover
models liberate only a tiny fraction of the false-vacuum
energy to radiation and have a relatively low reheat temperatures,
which is problematic for baryogenesis as it must proceed
after inflation.  The second problem lies in the name ``inflaton;'' because
the field responsible for inflation is so weakly coupled,
without heroic efforts it is difficult to make it an integral
part of a more encompassing particle physics theory.

\bigskip
\noindent{\bf First-order.}  These models have the potential
(no pun intended) to incorporate the best aspects of both
slow-rollover and old inflation.  Inflation is again
intimately connected to a cosmological phase transition
at a scale of order the GUT scale and no special
flatness is required of the Higgs potential.
Moreover, reheating proceeds via vacuum-bubble collisions which
guarantees good reheating and a unique signature of first-order
inflation, a background of gravitational waves proceeded
by bubble collisions, $\Omega_{\rm GW}\sim 10^{-8}$ at a frequency
determined by the scale of inflation, $f_{\rm GW}\sim
10^6\,{\rm Hz}\,({\cal M}/10^{12}\GeV )$ \cite{foigw}.

The simplest first-order inflation model is extended inflation.
First the good news:  Brans-Dicke gravity exhibits
conformal (scale) invariance (the Planck scale is replaced
by a field).  Conformal invariance is ``the Hallmark'' of
superstring theory, which has stimulated new interest in
Brans-Dicke like theories.  Now the bad news; in order to
avoid ``the big-bubble problem,'' the Brans-Dicke
parameter $\omega$ must be less than about 20, while solar-system
tests set a {\it lower limit} of about 500 \cite{solar}.  In
its simplest form, extended inflation is not viable.
Several variants have been put forth \cite{foi};
the simplest fix is to give the Brans-Dicke field a mass
\cite{kst}.  (A mass for the
Brans-Dicke field anchors at the right value and makes
the immune to solar-system tests.)  Any mass less than
about $10^9\GeV$ and greater than a tiny fraction of an eV
will do.  Moreover, this simple fix involves something
that string theorists must do anyway:  break conformal
invariance (the world is not conformally invariant,
it has a multitude of energy scales).

In sum, inflation provides a very attractive early
Universe paradigm.  Models of inflation are based
upon well defined, albeit very speculative,
physics at energy scales well below the Planck scale.
At present there is no standard model, or even a
particularly compelling model; there are, however, a variety
of models that work.  Given our general ignorance
about physics at energy scales $\gg 10^3\GeV$,
perhaps that should be enough for the time being.
In any case, while elegance, simplicity, and mathematical
beauty often provide guidance to the theorist,
in the end, experiment and observation are the
final arbiters.  As I will discuss toward the end,
observations involving structure formation are
starting to do just that.

\subsection{Initial Conditions:  No-hair theorems}

Inflation is cosmologically attractive because it
promises to account for our present nearly FRW space-time
starting from very general initial conditions.
Somewhat paradoxically, inflation is usually analyzed
in the context of the isotropic and homogeneous
FRW cosmology.  I will now explain the
{\it apparent} paradox and discuss to what extent inflation
lessens the dependence of the present state of the Universe
upon its initial state.

To begin consider the anisotropic but homogeneous
(Bianchi) models; the mean expansion rate
of the Universe can be written as
\begin{equation}
H^2 \equiv
({\dot{\overline R}}/{\overline R})^2 = {8\pi G\rho\over 3} +
F({\dot{\overline R}}, {\overline R});
\end{equation}
where ${\overline R}$ is the mean scale factor and
$\rho$ is the usual energy density and the function
$F$ accounts for the additional terms that arise due
to anisotropy.  In general,
the function $F$ decreases at least as rapidly as
$1/{\overline R}^2$, that is, as rapidly as the spatial
curvature term in the FRW cosmology or faster.  The false-vacuum
energy density appears in the energy density term
and is of course constant.  Provided that $F$ is positive,
the Universe will eventually become vacuum-energy
dominated; once it does, the $F({\dot{\overline R}}, {\overline R})$
term will quickly decrease and become insignificant
and the space time becomes isotropic.\footnote{There
is one worry; namely that the inflaton field will evolve to
the minimum of its potential before the vacuum-dominated phase
begins.  In general, this does not occur as anisotropy
increases the expansion rate, and thus the friction term
in the equation of motion for the inflaton; see \cite{TW}.}
This justifies the usual FRW analysis of inflation.

Not all anisotropic space-times will inflate; if
$F$ is sufficiently large and negative it will prevent inflation;
the simplest noninflating model is a very positively curved
FRW model that recollapses before it can inflate.
The strongly positively curved models
preclude a true cosmological no-hair theorem; however, it
has been shown that all spatially homogeneous,
but anisotropic models  eventually
inflate, except for the very positively curved models
\cite{nohair}.   And further, it has been shown
that ``smooth regions'' of inhomogeneous
models of sufficient size and that are negatively
curved will inflate \cite{js,aastar}.
While not all spacetimes will inflate, the class of spacetimes
that do is not special, but very generic \cite{aastar}.
Thus, inflation does indeed lessen
the dependence of the present state of the Universe on
its initial state.

Does inflation render a generic space-time isotropic and
homogeneous forever?  The answer is clearly no; the
most one can expect in an inhomogeneous space-time is
that negatively curved regions inflate.  Further, once
inflation is over, inhomogeneity and anisotropy will
``grow back.''  Consider spatial curvature; if the Universe
was not flat before inflation it will not be flat after
inflation.  However, inflation exponentially postpones
the epoch when spatial curvature becomes important because
the value of $\Omega$ after inflation becomes exponential
close to unity.  Likewise, in the exponential distant
future our Hubble volume will become larger than the generic
smooth patch created by inflation and we will in principle
we able to see the inhomogeneity beyond our inflationary patch \cite{frieman}.

Finally, there are the initial data for the scalar
field responsible for inflation itself.  In first-order
inflation, as in old inflation, this is a dynamical issue:
the initial value of the scalar field is determined by
thermal considerations.  However, in slow-rollover inflation
the story is very different; the initial value of the inflaton
field (and its spatial and temporal derivatives) are not
so determined, and at the classical level must be considered
to be initial data.  While this has become a subject unto
itself, some very general statements can be made.  First,
the inflaton field must be smooth on a scale comparable
to the Hubble radius, otherwise the energy density
associated with spatial gradients will dominate over the
vacuum energy preventing inflation.  Second, the value of
the scalar field must be small enough in models of ``new
inflation'' or large enough in models of ``chaotic inflation''
so that it takes the field more than 60 Hubble times to
roll to the bottom of the potential.   Finally, the initial
velocity of the inflaton (i.e., $\dot\phi$) must be small
enough so that it does not rapidly speed to the bottom of the potential.
For a given inflationary model, all of these considerations can be studied
and quantitative statements made about the necessary
initial data for the inflaton field \cite{piran}; further,
attempts have been made using the wavefunction of the Universe
to quantify the quantum expectation for the initial state
of the inflaton field \cite{lindeqm}.

In the final analysis it cannot be said that all initial
spacetimes undergo inflation and become isotropic and
homogeneous for all time; further, the initial data for the inflaton
itself must now be considered.
The strongest statement that one can make is to say
that inflation greatly lessens the dependence of the present state of
the Universe upon its initial state.  In my mind, that's
no mean feat and inflation should be considered a great success.

\section{Inflation:  The Fundamentals}

In this Section I discuss how to analyze an
inflationary model, given the scalar potential.
In two sections hence I will work through a
number of examples.  The focus will
be on the metric perturbations---density fluctuations \cite{scalar}
and gravity waves \cite{tensor}---that arise due to quantum fluctuations,
and the CBR temperature anisotropies that result
from them.\footnote{Isocurvature perturbations
can arise due to quantum fluctuations in other massless fields,
e.g., the axion field, if it exists \cite{isocurv}.}
Perturbations on all astrophysically interesting scales,
say $1\Mpc$ to $10^4\Mpc$, are produced during
an interval of about 8 e-folds around 50 e-folds
before the end of inflation, when these scales
crossed outside the horizon during inflation.
I will show how
the density perturbations and gravity waves
can be related to three features of the inflationary
potential:  its value $V_{50}$, its steepness
$x_{50}\equiv (\mpl V^\prime /V)_{50}$, and the change in
its steepness $x_{50}^\prime$, evaluated in
the region of the potential where the scalar
field was about 50 e-folds before the end of
inflation.  In principle, cosmological observations,
most importantly CBR anisotropy, can be used
to determine the characteristics of the
density perturbations and gravitational waves
and thereby $V_{50}$, $x_{50}$, and
$x_{50}^\prime$.

All viable models of inflation are of the
slow-rollover variety, or can be recast as
such \cite{allslow}.  In slow-rollover
inflation a scalar field
that is initially displaced from the minimum
of its potential rolls slowly to that minimum, and as it does
the cosmic-scale factor grows very rapidly.
Once the scalar field reaches the
minimum of the potential it oscillates about it, so that the large
potential energy has been converted into coherent
scalar-field oscillations, corresponding to
a condensate of nonrelativistic scalar particles.
The eventual decay of these particles into lighter
particle states and their subsequent thermalization
lead to the reheating of the Universe to a
temperature $T_{\rm RH} \simeq \sqrt{\Gamma \mpl}$,
where $\Gamma$ is the decay width of the
scalar particle \cite{reheat,allslow}.
Here, I will focus on the classical evolution of
the inflaton field during the slow-roll phase and
the small quantum fluctuations in the inflaton field which
give rise to density perturbations and those
in the metric which give rise to gravity waves.

To begin, let us assume that the scalar field driving inflation is
minimally coupled so that its stress-energy tensor
takes the canonical form,
\begin{equation}
T_{\mu\nu} = \partial_\mu\phi \partial_\nu\phi
-{\cal L}g_{\mu \nu};
\end{equation}
where the Lagrangian density of the scalar field
${\cal L} = {1\over 2}\partial_\mu\phi\partial^\mu\phi - V(\phi )$.
If we make the usual assumption that the scalar field
$\phi$ is spatially homogeneous, or at least so over a Hubble radius,
the stress-energy tensor takes the perfect-fluid form
with energy density, $\rho = {1\over 2}{\dot\phi}^2
+V(\phi )$, and isotropic pressure, $p={1\over 2}{\dot\phi}^2 -V(\phi )$.
The classical equations of motion
for $\phi$ can be obtained from the first law of thermodynamics,
$d(R^3\rho )=-pdR^3$, or by taking the four-divergence of $T^{\mu\nu}$:
\begin{equation}\label{eq:sr}
\ddot\phi + 3H\dot\phi + V^\prime (\phi ) = 0;
\end{equation}
the $\Gamma \dot\phi$ term responsible for reheating has
been omitted since we shall only be interested in the
slow-rollover phase.  In addition, there is the
Friedmann equation, which governs the expansion of the Universe,
\begin{equation}\label{eq:feq}
H^2 = {8\pi \over 3\mpl^2}\left( V(\phi ) + {1\over 2}
{\dot\phi}^2\right) \simeq {8\pi V(\phi )\over 3 \mpl^2};
\end{equation}
where we assume that the contribution of
all other forms of energy density, e.g., radiation and kinetic
energy of the scalar field, and
the curvature term ($k/R^2$) are negligible.
The justification for discussing inflation in the context
of a flat FRW model with a homogeneous scalar field
driving inflation were discussed earlier
(and at greater length in Ref.~\cite{inflation});
including the $\phi$ kinetic
term increases the righthand side of Eq. (\ref{eq:feq}) by a factor
of $(1+x^2/48\pi )$, a small correction for viable models.

In the next Section I will be more
precise about the amplitude of density perturbations and
gravitational waves; for now, let me briefly discuss how
these perturbations arise and give their characteristic amplitudes.
The metric perturbations produced in inflationary
models are very nearly ``scale invariant,''
a particularly simple spectrum which was first discussed
by Harrison and Zel'dovich \cite{hz}, and arise due
to quantum fluctuations.
In deSitter space all massless scalar fields experience
quantum fluctuations of amplitude $H/2\pi$.
The graviton is massless and can be described by
two massless scalar fields, $h_{+,\times} =
\sqrt{16\pi G}\phi_{+,\times}$ ($+$ and $\times$
are the two polarization states).  The inflaton by virtue of its flat
potential is for all practical purposes massless.

Fluctuations in the inflaton field lead to density fluctuations
because of its scalar potential, $\delta \rho \sim
HV^\prime$; as a given mode crosses outside the horizon, the density
perturbation on that scale becomes a classical metric
perturbation.  While outside the horizon, the description
of the evolution of a density perturbation is beset
with subtleties associated with the gauge freedom in
general relativity; there is, however, a simple gauge-invariant
quantity, $\zeta \simeq \delta\rho /(\rho +p)$, which
remains constant outside the horizon.  By equating the
value of $\zeta$ at postinflation horizon crossing with
its value as the scale crosses outside the horizon
it follows that $(\delta \rho /\rho )_{\rm HOR}
\sim HV^\prime /{\dot\phi}^2$ (note:  $\rho +p
= {\dot\phi}^2$); see Fig.~7.

The evolution of a gravity-wave perturbation is
even simpler; it obeys the massless Klein-Gordon
equation
\begin{equation}
{\ddot h}^i_k +3H{\dot h}^i_k + k^2h^i_k/R^2=0;
\end{equation}
where $k$ is the wavenumber of the mode and $i=+,\times$.  For
superhorizon sized modes, $k\la RH$, the solution
is simple:  $h^i_k=\,$const.  Like their density
perturbation counterparts, gravity-wave perturbations
become classical metric perturbations as they cross
outside the horizon; they are characterized by an
amplitude $h^i_k \simeq \sqrt{16\pi G}(H/2\pi )\sim
H/\mpl$.  At postinflation horizon crossing their
amplitude is unchanged.

Finally, let me write the horizon-crossing amplitudes
of the scalar and tensor metric perturbations
in terms of the inflationary potential,
\begin{eqnarray}\label{eq:scalartensor}
(\delta\rho /\rho)_{{\rm HOR},\lambda} & =  & c_S \left(
{V^{3/2}\over \mpl^3 V^\prime}\right)_1; \\
h_{{\rm HOR},\lambda} &  =  &
c_T \left({V^{1/2}\over \mpl^2}\right)_1 ;
\end{eqnarray}
where $(\delta\rho /\rho )_{{\rm HOR},\lambda}$ is the amplitude
of the density perturbation on the scale $\lambda$
when it crosses the Hubble radius during the post-inflation epoch,
$h_{{\rm HOR},\lambda}$ is the dimensionless amplitude
of the gravitational wave perturbation on the scale $\lambda$
when it crosses the Hubble radius, and $c_S$, $c_T$
are numerical constants of order unity.
Subscript 1 indicates that the quantity involving
the scalar potential is to be evaluated when the
scale in question crossed outside the
horizon during the inflationary era.  The metric
perturbations produced by inflation are characterized
by almost scale-invariant horizon-crossing amplitudes;
the slight deviations from scale invariance result
from the variation of $V$ and $V^\prime$ during inflation
which enter through the dependence upon $t_1$.
[In Eq. (\ref{eq:scalartensor})
I got ahead of myself and used the slow-roll
approximation (see below)
to rewrite the expression, $(\delta \rho /\rho )_{{\rm HOR},
\lambda}\simeq HV^\prime /\dot\phi$, in terms of the potential only.]

Eqs. (\ref{eq:sr}-\ref{eq:scalartensor}) are the
fundamental equations that govern
inflation and the production of metric perturbations.
It proves very useful to recast these equations using
the scalar field as the independent variable; we
then express the scalar and tensor perturbations
in terms of the value of the potential, its steepness,
and the rate of change of its steepness when the interesting
scales crossed outside the Hubble radius during inflation,
about 50 e-folds in scale factor before the end of inflation,
defined by
$$V_{50} \equiv V(\phi_{50}); \qquad
x_{50} \equiv {\mpl V^\prime (\phi_{50})\over
V(\phi_{50})}; \qquad x_{50}^\prime = {\mpl V^{\prime\prime}
(\phi_{50})\over V(\phi_{50})} - {\mpl [V^\prime (\phi_{50})]^2
\over V^2(\phi_{50})}.$$

To evaluate these three quantities 50 e-folds before
the end of inflation we must find the value of
the scalar field at this time.
During the inflationary phase the $\ddot\phi$ term is
negligible (the motion of $\phi$ is friction dominated),
and Eq. (\ref{eq:sr}) becomes
\begin{equation}\label{eq:sra}
\dot\phi \simeq {-V^\prime (\phi) \over 3H};
\end{equation}
this is known as the slow-roll approximation \cite{st}.
While the slow-roll approximation is almost
universally applicable, there are models where
the slow-roll approximation cannot be used;
e.g., a potential where during the crucial 8 e-folds
the scalar field rolls uphill, ``powered'' by the
velocity it had when it hit the incline.

The conditions that must be satisfied in order that
$\ddot\phi$ be negligible are:
\begin{eqnarray}\label{eq:condx}
|V^{\prime\prime}| & < & 9H^2 \simeq 24\pi V/\mpl^2; \\
|x| \equiv |V^\prime \mpl /V|  & <  & \sqrt{48\pi}.
\end{eqnarray}
The end of the slow roll occurs
when either or both of these inequalities
are saturated, at a value of $\phi$ denoted by $\phi_{\rm end}$.
Since $H\equiv {\dot R} /R$,
or $Hdt = d\ln R$, it follows that
\begin{equation}
d\ln R = {8\pi\over \mpl^2}\, {V(\phi ) d\phi
\over -V^\prime (\phi )} = -{8\pi d\phi \over \mpl \,x}.
\end{equation}
Now express the cosmic-scale factor
in terms of is value at the end of inflation, $R_{\rm end}$,
and the number of e-foldings before the end of inflation, $N(\phi )$,
$$R = \exp [-N(\phi )]\,R_{\rm end}. $$
The quantity $N(\phi )$ is a time-like variable whose
value at the end of inflation is zero and whose
evolution is governed by
\begin{equation}\label{eq:neq}
{dN \over d\phi} = {8\pi \over \mpl\,x} .
\end{equation}
Using Eq. (\ref{eq:neq}) we can compute the value of
the scalar field 50 e-folds before the end of inflation ($\equiv
\phi_{50}$); the values of
$V_{50}$, $x_{50}$, and $x_{50}^\prime$ follow directly.

As $\phi$ rolls down its potential during inflation its
energy density decreases,
and so the growth in the scale factor is not exponential.
By using the fact that the stress-energy of the scalar
field takes the perfect-fluid form, we can solve
for evolution of the cosmic-scale factor.  Recall,
for the equation of state $p=\gamma \rho$, the scale factor grows as
$R\propto t^q$, where $q= 2/3(1+\gamma )$.  Here,
\begin{eqnarray}\label{eq:eos}
\gamma & = & {{1\over 2}{\dot\phi}^2 - V \over
        {1\over 2}{\dot\phi}^2 +V} = { x^2-48\pi \over
        x^2 +48\pi };  \\
q & = & {1\over 3} + {16\pi \over x^2}.
\end{eqnarray}
Since the steepness of the potential can change during
inflation, $\gamma$ is not in general constant;
the power-law index $q$ is more
precisely the logarithmic rate of the change of the
logarithm of the scale factor, $q=d\ln R/d\ln t$.

When the steepness parameter is small,
corresponding to a very flat potential, $\gamma$ is close
to $-1$ and the scale factor grows as a very large
power of time.  To solve the horizon problem the
expansion must be ``superluminal'' (${\ddot R}>0$),
corresponding to $q>1$, which requires that $x^2< 24\pi$.
Since ${1\over 2}{\dot\phi}^2/V = x^2/48\pi$,
this implies that ${1\over 2}{\dot\phi}^2 /V(\phi ) < {1\over 2}$,
justifying neglect of the scalar-field kinetic energy
in computing the expansion rate for all but the
steepest potentials.  (In fact there are much stronger
constraints; the COBE DMR data imply that $n \ga 0.5$,
which restricts $x_{50}^2 \la 4\pi$, ${1\over 2}{\dot\phi}^2/V
\la {1\over 12}$, and $q\ga 4$.)

Next, let us relate the size of a given scale to
when that scale crosses outside
the Hubble radius during inflation, specified by $N_1(\lambda )$,
the number of e-folds before the end of inflation.
The physical size of a perturbation is related to its comoving size,
$\lambda_{\rm phys}=R\lambda$; with the usual
convention, $R_{\rm today} =1$, the comoving size
is the physical size today.  When the scale $\lambda$
crosses outside the Hubble radius $R_1
\lambda = H_1^{-1}$.  We then assume that:  (1) at the end of
inflation the energy density is ${\cal M}^4\simeq
V(\phi_{\rm end})$; (2) inflation is followed by a
period where the energy density of the Universe is dominated by coherent
scalar-field oscillations which decrease as
$R^{-3}$; and (3) when value of the scale factor
is $R_{\rm RH}$ the Universe reheats to a temperature
$T_{\rm RH} \simeq \sqrt{\mpl \Gamma}$ and expands
adiabatically thereafter.  The ``matching equation''
that relates $\lambda$ and $N_1(\lambda )$ is:
\begin{equation}\label{eq:prematch}
\lambda =  {R_{\rm today}\over R_1 } H_1^{-1}
        = {R_{\rm today} \over R_{\rm RH}} \,
        {R_{\rm RH}\over R_{\rm end}}\,
        {R_{\rm end}\over R_1}\, H_1^{-1}.
\end{equation}
Adiabatic expansion since reheating implies
$R_{\rm today}/R_{\rm RH} \simeq T_{\rm RH}/2.73\,{\rm K}$;
and the decay of the coherent scalar-field oscillations
implies $(R_{\rm RH}/R_{\rm end})^3 = ({\cal M}/T_{\rm RH})^4$.
If we define ${\bar q} = \ln (R_{\rm end}/R_1)/\ln (t_{\rm end}/t_1)$,
the mean power-law index, it follows that $(R_{\rm end}/R_1)H_1^{-1} =
\exp [N_1({\bar q}-1)/{\bar q}]H_{\rm end}^{-1}$,
and Eq. (\ref{eq:prematch}) becomes
\begin{equation}\label{eq:match}
N_1(\lambda )=  {{\bar q}\over {\bar q} -1}\,
\left[48+\ln \lambda_{\Mpc}
+ {2\over 3}\ln({\cal M}/10^{14}\GeV) + {1\over 3}
\ln (T_{\rm RH}/10^{14}\GeV ) \right];
\end{equation}
In the case of perfect reheating, which probably only applies to
first-order inflation, $T_{\rm RH} \simeq{\cal M}$.

The scales of astrophysical interest
today range roughly from that of galaxy size,
$\lambda \sim \Mpc$, to the present
Hubble scale, $H_0^{-1}\sim 10^4\Mpc$; up to the
logarithmic corrections these
scales crossed outside the horizon between about
$N_1(\lambda )\sim 48$ and $N_1(\lambda )\simeq 56$
e-folds before the end of inflation.  {\it That is,
the interval of inflation that determines its all observable
consequences covers only about 8 e-folds.}

Except in the case of strict power-law inflation
$q$ varies during inflation; this means that the
$(R_{\rm end}/R_1)H_1^{-1}$ factor in Eq. (\ref{eq:prematch})
cannot be written in closed form.
Taking account of this, the matching equation becomes
a differential equation,
\begin{equation}\label{eq:dmatch}
{d\ln\lambda_{\Mpc}\over dN_1} = {q(N_1) -1  \over q(N_1)};
\end{equation}
subject to the ``boundary condition:''
$$\ln \lambda_{\Mpc}
= -48 -{4\over 3}\ln ({\cal M}/10^{14}\GeV )+{1\over 3}\ln
(T_{\rm RH}/10^{14}\GeV )$$
for $N_1=0$, the matching relation
for the mode that crossed outside the Hubble radius at the
end of inflation.  Equation (\ref{eq:dmatch}) allows
one to obtain the precise expression for when a given scale
crossed outside the Hubble radius during inflation.  To
actually solve this equation, one would need to supplement it
with the expressions $dN/d\phi = 8\pi /\mpl x$ and
$q = 16\pi /x^2$.  For our purposes we need only know:  (1) The scales of
astrophysical interest correspond to $N_1\sim ``50\pm 4$,''
where for definiteness we will throughout take this
to be an equality sign.  (2) The expansion of Eq.
(\ref{eq:dmatch}) about $N_1 =50$,
\begin{equation}\label{eq:dndl}
\Delta N_1(\lambda ) = \left( {q_{50}-1\over q_{50} } \right)
\Delta \ln \lambda_{\Mpc} ;
\end{equation}
which, with the aid of Eq. (\ref{eq:neq}), implies that
\begin{equation} \label{eq:philambda}
\Delta \phi = \left( {q_{50} -1\over q_{50}}\right)\,
{x_{50}\over 8\pi}\, \Delta \lambda_{\Mpc} .
\end{equation}

We are now ready to express the perturbations
in terms of $V_{50}$, $x_{50}$, and
$x_{50}^\prime$.  First, we must solve for the value of $\phi$,
50 e-folds before the end of inflation.
To do so we use Eq. (\ref{eq:neq}),
\begin{equation}\label{eq:phi50}
N(\phi_{50} ) = 50 = {8\pi \over \mpl^2}\int_{\phi_{\rm end}}^{\phi_{50}}
{Vd\phi \over V^\prime}.
\end{equation}
Next, with the help of Eq. (\ref{eq:philambda})
we expand the potential $V$ and its steepness
$x$ about $\phi_{50}$:
\begin{equation}\label{eq:expandV}
V \simeq V_{50} + V_{50}^\prime (\phi - \phi_{50} )
= V_{50}\left[ 1 +  {x_{50}^2\over 8\pi}\,\left( {q_{50}\over q_{50}-1}
\right) \,  \Delta\ln \lambda_{\Mpc} \right] ;
\end{equation}
\begin{equation}\label{eq:expandx}
x \simeq x_{50} + x^\prime_{50}(\phi - \phi_{50})
= x_{50}\left[ 1 + {\mpl x_{50}^{\prime}
\over 8\pi}\,\left( {q_{50} \over q_{50}-1}\right)
\,\Delta\ln\lambda_{\Mpc} \right];
\end{equation}
of course these expansions only make sense for
potentials that are smooth.  We note that additional
terms in either expansion are ${\cal O}(\alpha_i^2)$ and
beyond the accuracy we are seeking.

Now recall the equations for the amplitude of the
scalar and tensor perturbations,
\begin{eqnarray}\label{eq:perts}
(\delta \rho /\rho )_{{\rm HOR},\lambda} & = & c_S \left( {V^{1/2}\over
\mpl^2 x}\right)_1 ;\\
h_{{\rm HOR},\lambda} & = & c_T\left( {V^{1/2}\over \mpl^2} \right)_1 ;
\end{eqnarray}
where subscript 1 means that the quantities are to be
evaluated where the scale $\lambda$ crossed outside the
Hubble radius, $N_1(\lambda )$ e-folds before the
end of inflation.   The origin of any deviation from
scale invariance is clear:  For tensor perturbations it
arises due to the variation of the potential; and for
scalar perturbations it arises due to the
variation of both the potential and its steepness.

Using Eqs. (\ref{eq:dndl}-\ref{eq:perts}) it is now
simple to calculate the power-law exponents $\alpha_S$
and $\alpha_T$ that quantify the deviations from scale
invariance,
\begin{eqnarray}\label{eq:alpha}
\alpha_T  &  =  & {x_{50}^2\over 16\pi}\,{q_{50}\over q_{50}-1} \simeq
        {x_{50}^2 \over 16\pi};  \\
\alpha_S &  =  &  \alpha_T - {\mpl x_{50}^\prime \over 8\pi}\,{q_{50}
\over q_{50} -1} \simeq {x_{50}^2\over 16\pi}
- {\mpl x_{50}^\prime \over 8\pi};
\end{eqnarray}
where
\begin{eqnarray}\label{eq:defofsi}
q_{50} &  =  &  {1\over 3} + {16\pi \over x_{50}^2}
        \simeq {16\pi \over x_{50}^2} ;\\
h_{{\rm HOR},\lambda} & = & c_T \left( {V_{50}^{1/2}\over
\mpl^2 } \right) \, \lambda_{\Mpc}^{\ \ \,\alpha_T};\\
(\delta\rho /\rho )_{{\rm HOR},\lambda} &  =  &  c_S
\left({V_{50}^{1/2}\over x_{50}\mpl^2}\right)\,
\lambda_{\Mpc}^{\ \ \,\alpha_S}.
\end{eqnarray}
The spectral indices $\alpha_i$ are defined as,
$\alpha_S = [d\ln (\delta\rho /\rho)_{{\rm HOR},
\lambda} / d\ln \lambda_{\Mpc}]_{50}$ and $\alpha_T =
[d\ln h_{{\rm HOR},\lambda} /d\ln
\lambda_{\Mpc}]_{50}$, and in general vary
slowly with scale.  Note too that the
deviations from scale invariance, quantified
by $\alpha_S$ and $\alpha_T$, are of the order of
$x_{50}^2$, $\mpl x_{50}^\prime$.  In the
expressions above we retained only lowest-order
terms in ${\cal O}(\alpha_i)$.  The next-order contributions to the
spectral indices are ${\cal O}(\alpha_i^2)$; those
to the amplitudes are ${\cal O}(\alpha_i)$ and
are given two sections hence.  The justification for
truncating the expansion at lowest order is that
the deviations from scale invariance are expected to be small---and
are required by astrophysically data to be small.

As I discuss in more detail two sections hence, the more
intuitive power-law indices $\alpha_S$, $\alpha_T$ are
related to the indices that are usually used
to describe the power spectra of scalar and tensor
perturbations, $P_S(k) = |\delta_k|^2 = A k^n$
and $P_T(k) = |h_k|^2 = A_Tk^{n_T}$,
\begin{eqnarray}\label{eq:indices}
n & = & 1-2\alpha_S = 1 -{x_{50}^2\over 8\pi}
+ {\mpl x_{50}^\prime\over 4\pi} ; \\
n_T & = & -2\alpha_T =  -{x_{50}^2\over 8\pi} .\\
\end{eqnarray}

CBR temperature fluctuations on large-angular scales
($\theta \ga 1^\circ$) due to metric perturbations
arise through the Sachs-Wolfe effect; very roughly, the
temperature fluctuation on a given angular scale
$\theta$ is related to the metric fluctuation on the
length scale that subtends that angle at last scattering,
$\lambda \sim 100h^{-1}\Mpc (\theta /{\rm deg})$,
\begin{eqnarray}
\left({\delta T\over T}\right)_\theta & \sim &
\left({\delta \rho \over \rho}\right)_{{\rm HOR},\lambda};\\
\left({\delta T\over T}\right)_\theta & \sim &
h_{{\rm HOR},\lambda} ;
\end{eqnarray}
where the scalar and tensor contributions to the CBR
temperature anisotropy on a given scale add in quadrature.
Let me be more specific about the amplitude of the
quadrupole CBR anisotropy.   For small $\alpha_S$,
$\alpha_T$ the contributions of each to
the quadrupole CBR temperature anisotropy:
\begin{eqnarray}\label{eq:quadanisotropy}
\left({\Delta T\over T_0}\right)_{Q-S}^2  &  \approx  &
{32\pi\over 45}{V_{50}\over \mpl^4 x_{50}^2};\\
\left({\Delta T\over T_0}\right)_{Q-T}^2  & \approx  &
0.61{V_{50}\over \mpl^4};
\end{eqnarray}
\begin{equation}\label{eq:ratio}
{T\over S}  \equiv {(\Delta T/T_0)_{Q-T}^2 \over
(\Delta T/T_0)_{Q-S}^2} \approx {0.28  x_{50}^2};
\end{equation}
where expressions have been evaluated
to lowest order in $x_{50}^2$ and $\mpl x_{50}^\prime$.
These quantities represent the ensemble averages of the
scalar and tensor contributions to the
quadrupole temperature anisotropy, which in terms of
the spherical-harmonic expansion of the CBR temperature anisotropy
on the sky are given by $5\langle |a_{2m}|^2\rangle /4\pi$.  Further,
the scalar and tensor contributions
to the {\it measured} quadrupole anisotropy add
in quadrature, and are subject
to ``cosmic variance.''   (Cosmic
variance refers to the dispersion in the values measured
by different observers in the Universe.)

Before going on, some general
remarks \cite{turner}.  The steepness parameter $x_{50}^2$ must
be less than about $24\pi$ to ensure
superluminal expansion.   For ``steep'' potentials,
the expansion rate is ``slow,'' i.e., $q_{50}$ closer to unity,
the gravity-wave contribution to the quadrupole CBR temperature anisotropy
becomes comparable to, or greater than, that of density
perturbations, and both scalar and tensor
perturbations exhibit significant deviations from scale invariance.
For ``flat'' potentials, i.e., small $x_{50}$,
the expansion rate is ``fast,'' i.e., $q_{50} \gg 1$,
the gravity-wave contribution to the quadrupole CBR temperature anisotropy
is much smaller than that of density perturbations, and
the tensor perturbations are scale invariant.  Unless
the steepness of the potential changes rapidly, i.e.,
large $x_{50}^\prime$, the scalar perturbations are also
scale invariant.

\subsection{Metric perturbations and CBR anisotropy}

I was purposefully vague when discussing
the amplitudes of the scalar and tensor modes, except
when specifying their contributions to the quadrupole CBR temperature
anisotropy; in fact, the spectral indices $\alpha_S$ and
$\alpha_T$, together with the scalar and tensor
contributions to the CBR quadrupole serve to
provide all the information necessary.  Here I will
fill in more details about the metric perturbations.

The scalar and tensor metric perturbations are expanded
in harmonic functions, in the flat Universe predicted
by inflation, plane waves,
\begin{eqnarray}
h_{\mu\nu}({\bf x}, t) & = & {1\over (2\pi )^3}
\int d^3k\,h_{\bf k}^i (t)\, \varepsilon_{\mu\nu}^i
\, e^{-i{\bf k}\cdot{\bf x}} ;\\
{\delta\rho ({\bf x},t) \over \rho} & = &
{1\over (2\pi )^3}\int d^3k\,\delta_{\bf k} (t) \, e^{-i{\bf k}\cdot{\bf x}} ;
\end{eqnarray}
where $h_{\mu\nu}=R^{-2}g_{\mu\nu} - \eta_{\mu\nu}$,
$\varepsilon_{\mu\nu}^i$ is the polarization tensor
for the gravity-wave modes, and $i= +$, $\times$ are
the two polarization states.  Everything of interest
can be computed in terms of $h_{\bf k}^i$ and $\delta_{\bf k}$.
For example, the {\it rms} mass fluctuation
in a sphere of radius $r$ is obtained in terms of the
window function for a sphere and the power spectrum $P_S(k)
\equiv \langle |\delta_{\bf k}|^2\rangle$ (see below),
\begin{equation}
\langle (\delta M /M)^2\rangle_r = {9\over 2\pi^2r^2}\,
\int_0^\infty [j_1(kr)]^2 \,P_S(k) dk ;
\end{equation}
where $j_1(x)$ is the spherical Bessel function of first order.
If $P_S(k)$ is a power law, it follows roughly that
$(\delta M/M)^2 \sim k^3|\delta_{\bf k}|^2$,
evaluated on the scale $k=r^{-1}$.
This is what I meant by $(\delta \rho /
\rho)_{{\rm HOR},\lambda}$:  the {\it rms} mass
fluctuation on the scale $\lambda$
when it crossed inside the horizon.  Likewise,
by $h_{{\rm HOR},\lambda}$ I meant the {\it rms} strain
on the scale $\lambda$ as it crossed inside the Hubble radius,
$(h_{{\rm HOR},\lambda})^2 \sim k^3|h_{\bf k}^i|^2$.

In the previous discussions I have chosen to specify
the metric perturbations for the different Fourier
modes when they crossed inside the horizon,
rather than at a common time.  I did so because
scale invariance is made manifest, as the scale independence
of the metric perturbations at post-inflation horizon crossing.
Recall, in the case of scalar perturbations
$(\delta \rho /\rho )_{\rm HOR}$ is up to a numerical factor
the fluctuation in the Newtonian potential, and, by specifying
the scalar perturbations at horizon crossing, we avoid the
discussion of scalar perturbations on superhorizon
scales, which is beset by the subtleties associated with
the gauge noninvariance of $\delta_{\bf k}$.

It is, however, necessary to specify the perturbations at a common time
to carry out most calculations; e.g., an $N$-body simulation
of structure formation or the calculation of CBR anisotropy.
To do so, one has to
take account of the evolution of the perturbations
after they enter the horizon.
After entering the horizon tensor perturbations behave like
gravitons, with $h_{\bf k}$ decreasing as $R^{-1}$ and
the energy density associated with a given mode, $\rho_k \sim
\mpl^2 k^5|h_{\bf k}|^2/R^2$, decreasing as $R^{-4}$.  The evolution
of scalar perturbations is slightly more complicated; modes that
enter the horizon while the Universe is still radiation dominated
remain essentially constant until the Universe becomes matter
dominated (growing only logarithmically);
modes that enter the horizon after the Universe becomes
matter dominated grow as the scale factor.
(The gauge noninvariance of $\delta_{\bf k}$ is not an important
issue for subhorizon size modes; here a Newtonian analysis
suffices, and there is only one growing mode, corresponding to
a density perturbation.)

The method for characterizing the scalar perturbations
is by now standard:  The spectrum of
perturbations is specified at the present
epoch (assuming linear growth for all scales); the spectrum at earlier
epochs can be obtained by multiplying $\delta_{\bf k}$
by $R(t)/R_{\rm today}$.  The inflationary metric perturbations
are gaussian;
thus $\delta_{\bf k}$ is a gaussian, random variable.  Its
statistical expectation value is
\begin{equation}
\langle \delta_{\bf k}\,\delta_{\bf q}\rangle
= P_S(k)(2\pi )^3 \delta^{(3)} ({\bf k} -{\bf q});
\end{equation}
where the power spectrum today is written as
\begin{equation}
P_S(k) \equiv Ak^n T(k)^2;
\end{equation}
$n=1-2\alpha_S$ ($=1$ for scale-invariant perturbations), and
$T(k)$ is the ``transfer function'' which encodes
the information about the post-horizon crossing evolution
of each mode and depends
upon the matter content of the Universe, e.g., baryons plus cold dark
matter, baryons plus hot dark matter, baryons plus
hot and cold dark matter, and so on.
The transfer function is defined so that $T(k)\rightarrow 1$
for $k\rightarrow 0$ (long-wavelength perturbations); an
analytic approximation to the cold dark matter transfer
function is given by \cite{stat}
\begin{equation}\label{eq:cdmtf}
T(k)   =  {\ln (1+2.34 q)/2.34q \over \left[ 1 + (3.89q) +(16.1q)^2
+ (5.46q)^3 + (6.71q)^4 \right]^{1/4}} ;
\end{equation}
where $q= k/(\Omega_0 h^2 \Mpc^{-1})$.  Inflationary
power spectra for different dark matter possibilities
are shown in Fig.~9.

The overall normalization factor
\begin{equation}\label{eq:scalarnorm}
A = {1024\pi^3 \over 75H_0^{3+n}}\,{V_{50} \over \mpl^4 x_{50}^2}
\,{[1+{7\over 6}n_T - {1\over 3}(n-1)]
\left\{ \Gamma [{3\over 2} -{1\over 2}(n-1)]\right\}^2
\over 2^{n-1} [\Gamma ({3\over 2})]^2}\,k_{50}^{1-n} ;
\end{equation}
where the ${\cal O}(\alpha_i)$ correction to $A$
has been included \cite{ls}.  The quantity $n_T =-2\alpha_T =
-x_{50}^2/8\pi$, $n-1 = -2\alpha_S = n_T +x_{50}^\prime /4\pi$,
$k_{50}$ is the comoving wavenumber of the scale
that crossed outside the horizon 50 e-folds before the
end of inflation.  All the formulas below simplify if this
scale corresponds to the present horizon scale,
specifically, $k_{50}=H_0/2$.  [Eq. (\ref{eq:scalarnorm})
can be simplified by expanding $\Gamma ({3\over 2} + x )
= \Gamma (3/2)[1+x
(2-2\ln 2 -\gamma)]$, valid for $|x|\ll 1$;
$\gamma \simeq 0.577$ is Euler's constant.]

\begin{figure}
\vspace{4.6in}
\caption[power]{Power spectra for cold dark matter (CDM),
hot dark matter (HDM), mixed dark matter (MDM = 30\% hot +
70\% cold), and cold dark matter
with a cosmological constant ($\Lambda$CDM = 20\% CDM +
80\% $\Lambda$).  All spectra are normalized
to the COBE DMR quadrupole temperature anisotropy; $h=0.5$
for all models except $\Lambda$CDM ($h=0.8$).}
\end{figure}

{}From this expression it is simple to compute the
Sachs-Wolfe contribution of scalar perturbations
to the CBR temperature anisotropy; on angular scales much greater
than about $1^\circ$ (corresponding to multipoles $l \ll 100$) it
is the dominant contribution.  If we expand the
CBR temperature on the sky in spherical harmonics,
\begin{equation}
{\delta T(\theta ,\phi )\over T_0} = \sum_{l\ge 2, m=-l}^{l=\infty ,m=l}
a_{lm}Y_{lm}(\theta ,\phi );
\end{equation}
where $T_0=2.73\,$K is the CBR temperature today, then
the ensemble expectation for the multipole coefficients
is given by
\begin{eqnarray}\label{eq:scalarlm}
\langle |a_{lm}|^2\rangle & = & {H_0^4\over 2\pi}\,
\int_0^\infty k^{-2}\,P_S(k)\,|j_l(k r_0)|^2\,dk ; \\
&  \simeq &  {AH_0^{3+n}\, r_0^{1-n}\over 16}\,
{\Gamma (l+{1\over 2}n-{1\over 2})
\Gamma (3-n) \over \Gamma (l-{1\over 2}n +{5\over 2})
[\Gamma (2-{1\over 2}n)]^2} ;
\end{eqnarray}
where $r_0\approx 2H_0^{-1}$ is the comoving distance to the last scattering
surface, and this expression is for the Sachs-Wolfe contribution from scalar
perturbations only.  For $n$ not too different from
one, the ensemble expectation for the quadrupole CBR temperature
anisotropy is
\begin{equation}
\left( {\Delta T\over T_0} \right)_{Q-S}^2 \equiv
{5 |a_{2m}|^2 \over 4\pi } \approx {32\pi\over 45}\,
{V_{50} \over \mpl^4\,x_{50}^2}\,(k_{50}r_0)^{1-n}.
\end{equation}
(By choosing $k_{50}= r_0^{-1}= {1\over 2}H_0$,
the last factor becomes unity.)

The ensemble expectation values for the multipole amplitudes
are often referred to as the angular power spectrum.
Further, the {\it rms} temperature fluctuation on a
given angular scale is related to the multipole amplitudes
\begin{equation}
\left( {\Delta T\over T}\right)^2_\theta \sim
l^2\langle |a_{lm}|^2\rangle \qquad {\rm for}\ \ l \simeq 200^\circ /\theta .
\end{equation}

The procedure for specifying the tensor modes is similar,
cf. Refs.~\cite{aw,white}.  For the modes that enter the
horizon after the Universe becomes matter dominated,
$k\la 0.1h^2\Mpc$, which are the only modes that contribute
significantly to CBR anisotropy on angular scales
greater than a degree,
\begin{equation}
h_{\bf k}^i (\tau ) = a^i ({\bf k}) \left( { 3j_1(k\tau )\over
k\tau }  \right) ;
\end{equation}
where $\tau = r_0(t/t_0)^{1/3}$ is conformal time.
[For the modes that enter the horizon during the radiation-dominated
era, $k \ga 0.1h^2\Mpc^{-1}$, the factor
$3j_1(k\tau )/k\tau$ is replaced by $j_0(k\tau )$
for the remainder of the radiation era.
In either case, the factor involving the spherical Bessel
function quantifies the fact that tensor perturbations
remain constant while outside the horizon, and after
horizon crossing decrease as $R^{-1}$.]

The tensor perturbations too are characterized by
a gaussian, random variable, here written as $a^i({\bf k})$;
the statistical expectation
\begin{equation}
\langle h_{\bf k}^i h_{\bf q}^j \rangle =
P_T (k)(2\pi )^6 \delta^{(3)} ({\bf k} -{\bf q})\delta_{ij};
\end{equation}
where the power spectrum
\begin{eqnarray}
P_T(k) & = & A_T k^{n_T -3} \left[ {3j_1(k\tau )\over k\tau}\right]^2 ; \\
A_T & = & {8 \over 3\pi }\, {V_{50} \over \mpl^4}\,
{(1+{5\over 6}n_T)[\Gamma ({3\over 2} - {1\over 2}n_T)]^2 \over
2^{n_T} [\Gamma ({3\over 2})]^2}\, k_{50}^{-n_T} ;
\end{eqnarray}
where the ${\cal O}(\alpha_i)$ correction to $A_T$ has been
included.  Note that $n_T = -2\alpha_T$ is zero
for scale-invariant perturbations.

Finally, the contribution of tensor perturbations to
the multipole amplitudes, which arise solely due
to the Sachs-Wolfe effect \cite{SW,aw,white}, is given by
\begin{equation}\label{eq:tensorlm}
\langle |a_{lm}|^2 \rangle \simeq 36 \pi^2 \,{\Gamma (l+3)
\over \Gamma (l-1) }\, \int_0^\infty\,
k^{n_T+1}\,A_T \, |F_l(k)|^2\,dk ;
\end{equation}
where
\begin{equation}
F_l(k)  =  - \int_{r_D}^{r_0} \, dr \,
{j_2(kr)\over kr}\,\left[ {j_l(kr_0-kr) \over
(kr_0-kr)^2}\right] ;
\end{equation}
and $r_D = r_0/(1+z_D)^{1/2} \approx r_0/35$ is
the comoving distance to the horizon at
decoupling (= conformal time at decoupling).
Equation (\ref{eq:tensorlm}) is approximate
in that very short wavelength modes, $kr_0\gg 100$,
that crossed inside the horizon before matter-radiation
equality have not been properly taken into account; to
take them into account, the integrand must be multiplied
by a transfer function,
\begin{equation}
T(k) \simeq 1.0 + 1.44(k/k_{\rm EQ}) +2.54 (k/k_{\rm EQ})^2;
\end{equation}
where $k_{\rm EQ} \equiv H_0/(2\sqrt{2}-2)R_{\rm EQ}^{1/2}$
is the scale that entered the horizon at matter radiation
equality \cite{turner}.  In addition, for $l\ga 1000$, the
finite thickness of the last-scattering surface
must be taken into account.

The tensor contribution to the quadrupole CBR temperature
anisotropy for $n_T$ not too different from zero is
\begin{equation}
\left( {\Delta T\over T_0} \right)_{Q-T}^2 \equiv
{5|a_{2m}|^2\over 4\pi} \simeq 0.61 {V_{50}\over \mpl^4}\, (k_{50}r_0)^{-n_T};
\end{equation}
where the integrals in the previous expressions have been evaluated
numerically.

Both the scalar and tensor contributions to a given
multipole are dominated by wavenumbers $kr_0\sim l$.
For scale-invariant perturbations and small $l$,
both the scalar and tensor contributions to
$(l+{1\over 2})^2\langle |a_{lm}|^2\rangle$ are approximately constant.
The contribution of scalar perturbations to
$(l+{1\over 2})^2\langle |a_{lm}|^2\rangle$ begins
to decrease for $l\sim 150$ because the scalar
contribution to these multipoles is
dominated by modes that entered the horizon before matter
domination (and hence are suppressed by the
transfer function).  The contribution of tensor modes to
$(l+{1\over 2})^2\langle |a_{lm}|^2\rangle$ begins
to decrease for $l\sim 30$ because the tensor contribution
to these multipoles is
dominated by modes that entered the horizon before decoupling
(and hence decayed as $R^{-1}$ until decoupling).
Figure 10 shows the contribution of scalar and tensor
perturbations to the CBR anisotropy multipole amplitudes
(and includes both the tensor and scalar transfer functions);
the expected variance in the CBR multipoles is given by
the sum of the scalar and tensor contributions.

\begin{figure}
\vspace{4.5in}
\caption[alm]{Scalar and tensor contributions to the
CBR multipole moments:  $l(l+1)\langle |a_{lm}|^2\rangle/6
\langle |a_{2m}|^2\rangle$ for the scalar and $l(l+{1\over 2})
\langle |a_{lm}|^2\rangle /5\langle |a_{2m}|^2\rangle$ for
the tensor.  The tensor contribution begins to fall off
for $l\sim 30$; here $n-1 = n_T =0$, $z_{\rm DEC} = 1000$,
and $h=0.5$ (from \cite{tl}).}
\end{figure}

\subsection {Worked examples}

In this Section I apply the formalism developed
in the two previous sections to four specific models.
So that I can, where appropriate, solve numerically
for model parameters, I will:  (1) Assume that
the astrophysically interesting scales crossed
outside the horizon 50 e-folds before the end
of inflation; and (2) Use the COBE DMR quadrupole measurement,
$\langle (\Delta T )_{Q}^2 \rangle^{1/2}
\approx 16\pm 2\mu$K \cite{DMR}, to normalize
the scalar perturbations; using Eq. (\ref{eq:quadanisotropy}) this implies
\begin{equation}
V_{50} \approx 1.6\times 10^{-11} \,\mpl^4\,x_{50}^2 .
\end{equation}
Of course it is entirely possible that
a significant portion of the quadrupole anisotropy is
due to tensor-mode perturbations, in which case this normalization
must be reduced by a factor of $(1+T/S)^{-1}$.
And, it is straightforward
to change ``50'' to the number appropriate to a
specific model, or to normalize the perturbations another way.

Before going on let us use the COBE DMR quadrupole anisotropy to bound the
tensor contribution to the quadrupole anisotropy
and thereby the energy density that drives inflation:
\begin{equation}
V_{50}\la 6\times 10^{-11}\mpl^4.
\end{equation}
Thus, the tensor contribution to the CBR quadrupole
implies that the vacuum energy that drives inflation
must be much less than the Planck energy density,
strongly suggesting that inflation is not a
quantum-gravitational phenomenon.

\subsubsection {Exponential potentials}

There are a class of models that can be described in terms
of an exponential potential,
\begin{equation}
V(\phi ) = V_0\exp (-\beta\phi /\mpl ).
\end{equation}
This type of potential was first invoked in the
context of power-law inflation \cite{powerlaw}, and has
recently received renewed interest in the context
of extended inflation \cite{extended}.  In the simplest
model of extended, or first-order, inflation,
that based upon the Brans-Dicke-Jordan theory
of gravity \cite{extended}, $\beta$ is related to the Brans-Dicke
parameter:  $\beta^2 = 64\pi /(2\omega +3)$.

For such a potential the slow-roll conditions are
satisfied provided that $\beta^2 \la 24\pi$;
thus inflation does not end until
the potential changes shape, or in the case of
extended inflation, until the phase transition
takes place.  In either case we can relate
$\phi_{50}$ to $\phi_{\rm end}$,
\begin{equation}
N(\phi_{50} ) = 50 = {8\pi\over \mpl^2}\int_{\phi_{50}}^{\phi_{\rm end}}
{Vd\phi \over -V^\prime};\qquad \Rightarrow\ \
\phi_{50} = \phi_{\rm end} - 50\beta / 8\pi .
\end{equation}
Since $\phi_{\rm end}$ is in effect arbitrary,
the overall normalization of the potential is
irrelevant.  The two other parameters,
$x_{50}$ and $x_{50}^\prime$, are easy to compute:
\begin{equation}
x_{50} = -\beta ; \qquad x_{50}^\prime = 0.
\end{equation}
Using the COBE DMR normalization, we can relate
$V_{50}$ and $\beta$:
\begin{equation}
V_{50} = 1.6 \times 10^{-11}\,\mpl^4 \beta^2.
\end{equation}
Further, we can compute $q$, $\alpha_S$, $\alpha_T$,
and $T/S$:
\begin{equation}
q = 16\pi /\beta^2; \qquad T/S = 0.28\beta^2;\qquad
\alpha_T =\alpha_S = 1/(q-1)\simeq \beta^2/16\pi .
\end{equation}
Note, for the exponential potential, $q$, $\alpha_T=
\alpha_S$ are independent of epoch.  In the case of
extended inflation, $\alpha_S =\alpha_T =4/(2\omega +3)$;
since $\omega$ must be less than about 20 \cite{bigbubbles},
this implies significant tilt:  $\alpha_S=\alpha_T \ga 0.1$.

\subsubsection{Chaotic inflation}

These models are based upon a very simple potential:
\begin{equation}
V(\phi ) = a\phi^b;
\end{equation}
$b=4$ corresponds to Linde's original model
of chaotic inflation and $a$ is dimensionless \cite{chaotic},
and $b=2$ is a model based upon a massive scalar field
and $m^2 = 2a$ \cite{massive}.  In these models $\phi$ is
initially displaced from $\phi = 0$, and inflation
occurs as $\phi$ slowly rolls to the origin.
The value of $\phi_{\rm end}$
is easily found:  $\phi_{\rm end}^2 = b(b-1)\mpl^2/24\pi$, and
\begin{eqnarray}
N(\phi_{50})=50 & =  & {8\pi \over \mpl^2}\int_{\phi_{\rm end}}^{\phi_{50}}
{Vd\phi \over V^\prime} ;\\
& \Rightarrow & \ \
\phi_{50}^2/\mpl^2  =  50b/4\pi + b^2/48\pi \simeq 50b/4\pi ;
\end{eqnarray}
the value of $\phi_{50}$ is a few times the Planck mass.

For purposes of illustration consider $b=4$; $\phi_{\rm end}
=\mpl /\sqrt{2\pi} \simeq 0.4\mpl$, $\phi_{50} \simeq
4\mpl$, $\phi_{46} \simeq 3.84\mpl$,
and $\phi_{54}\simeq 4.16\mpl$.  In order to have sufficient
inflation the initial value of $\phi$ must exceed about
$4.2\mpl$; inflation ends when $\phi \approx 0.4\mpl$; and
the scales of astrophysical interest cross outside the horizon
over an interval $\Delta \phi \simeq 0.3\mpl$.

The values of the potential, its steepness, and the
change in steepness are easily found,
\begin{equation}
V_{50} = a\,\mpl^b\,\left({50b\over 4\pi}\right)^{b/2};
\qquad x_{50} = \sqrt{4\pi b\over 50}; \qquad
\mpl x_{50}^\prime = {-4\pi \over 50} ;
\end{equation}
\begin{equation}
q_{50} = 200/b; \qquad
T/S = 0.07b; \qquad \alpha_T\simeq b/200; \qquad
\alpha_S = \alpha_T + 0.01 .
\end{equation}
Unless $b$ is very large, scalar perturbations dominate
tensor perturbations \cite{star}, $\alpha_T$, $\alpha_S$ are very
small, and $q$ is very large.  Further, when $\alpha_T$,
$\alpha_S$ become significant, they are equal.
Using the COBE DMR normalization we find:
\begin{equation}
a = 1.6\times 10^{-11} b^{1-b/2} (4\pi /50)^{b/2+1}\,\mpl^{4-b}.
\end{equation}
For the two special cases of interest: $b=4$, $a=6.4\times 10^{-14}$;
and $b=2$, $m^2 \equiv 2a = 2.0\times 10^{-12}\mpl^2$.

\subsubsection {New inflation}

These models entail
a very flat potential where the scalar field rolls from
$\phi \approx 0$ to the minimum of the potential at $\phi
=\sigma$.  The original models of slow-rollover inflation \cite{new}
were based upon potentials of the Coleman-Weinberg form
\begin{equation}
V(\phi ) = B\sigma^4/2 +B\phi^4\left[ \ln (\phi^2/\sigma^2)
-{1\over 2} \right] ;
\end{equation}
where $B$ is a very small dimensionless
coupling constant.  Other very flat potentials also work (e.g.,
$V = V_0 - \alpha\phi^4 +\beta \phi^6$ \cite{st}).  As before
we first solve for $\phi_{50}$:
\begin{equation}
N(\phi_{50}) = 50 = {8\pi\over \mpl^2}\int_{\phi_{\rm end}}^{\phi_{50}}
{Vd\phi\over V^\prime};\qquad \Rightarrow \ \
\phi_{50}^2 = {\pi \sigma^4\over 100 |\ln (\phi_{50}^2/\sigma^2)|\mpl^2};
\end{equation}
where the precise value of $\phi_{\rm end}$ is not relevant,
only the fact that it is much larger than $\phi_{50}$.
Provided that $\sigma \la \mpl$, both $\phi_{50}$ and
$\phi_{\rm end}$ are much less than $\sigma$; we then find
\begin{equation}
V_{50}\simeq B\sigma^4/2; \qquad x_{50} \simeq - {(\pi /25)^{3/2}
\over \sqrt{|\ln (\phi_{50}^2/\sigma^2|) }} \left( {\sigma\over \mpl}
\right)^2 \ll 1;
\end{equation}
\begin{equation}
\mpl x_{50}^\prime \simeq -24\pi /100;\qquad
q_{50} \simeq {2.5\times 10^5 |\ln (\phi_{50}^2/\sigma^2)|\over \pi^2}
\left({\mpl\over \sigma}\right)^4 \gg 1;
\end{equation}
\begin{equation}
\alpha_S \simeq {1\over q_{50}} \ll 1;\qquad
\alpha_T = \alpha_S + 0.03;\qquad
{T\over S} \simeq {6\times 10^{-4}\over |\ln (\phi_{50}^2/\sigma^2)|}\left(
{\sigma\over \mpl}\right)^4.
\end{equation}
Provided that $\sigma \la \mpl$, $x_{50}$ is very small; this means that
$q$ is very large, gravity-waves
and density perturbations are very nearly scale invariant,
and $T/S$ is small.  Finally, using the COBE DMR normalization,
we can determine the dimensionless coupling constant $B$:
\begin{equation}
B \simeq 6\times 10^{-14}/|\ln (\phi_{50}^2/\sigma^2)|
\approx 3\times 10^{-15}.
\end{equation}

\subsubsection{Natural inflation}

This model is based upon a potential of the form \cite{natural}
\begin{equation}
V(\phi ) = \Lambda^4 \left[ 1+\cos (\phi /f ) \right].
\end{equation}
The flatness of the potential (and requisite small
couplings) arise because the $\phi$ particle is a pseudo-Nambu-Goldstone
boson ($f$ is the scale of spontaneous symmetry breaking
and $\Lambda$ is the scale of explicit symmetry breaking; in
the limit that $\Lambda \rightarrow 0$ the $\phi$ particle
is a massless Nambu-Goldstone boson).   It is a simple
matter to show that $\phi_{\rm end}$ is of the order of $\pi f$.

This potential is difficult to analyze in
general; however, there are two limiting regimes:
(i) $f\gg \mpl$; and (ii) $f\la \mpl$ \cite{st}.  In the first
regime, the 50 or so relevant e-folds take place close
to the minimum of the potential, $\sigma = \pi f$, and
inflation can be analyzed by expanding the potential about $\phi=\sigma$,
\begin{equation}
V(\psi ) \simeq {m^2\psi^2}/2 ;
\end{equation}
\begin{equation}
m^2 = \Lambda^4 /f^2; \qquad \psi = \phi -\sigma .
\end{equation}
In this regime natural inflation is equivalent to chaotic
inflation with $m^2 =\Lambda^4 /f^2 \simeq 2\times 10^{-12}\mpl^2$.

In the second regime, $f\la \mpl$, inflation takes
place when $\phi \la \pi f$, so that we can make the
following approximations:  $V \simeq 2\Lambda^4$
and $V^\prime = - \Lambda^4\phi /f^2$.  Taking
$\phi_{\rm end} \sim \pi f$, we can solve for $N(\phi )$:
\begin{equation}
N(\phi ) = {8\pi \over \mpl^2}\int_\phi^{\pi f} {Vd\phi \over -V^\prime}
\simeq {16\pi \mpl^2\over f^2} \ln (\pi f/\phi ) ;
\end{equation}
from which it is clear that achieving 50 e-folds of
inflation places a lower bound to $f$, very
roughly $f\ga \mpl /3$ \cite{st,natural}.

Now we can solve for $\phi_{50}$, $V_{50}$, $x_{50}$,
and $x_{50}^\prime$:
\begin{equation}
\phi_{50}/\pi f \simeq \exp (-50 \mpl^2/16\pi f^2)\la {\cal O}(0.1);
\qquad V_{50} \simeq 2\Lambda^4;
\end{equation}
\begin{equation}
x_{50} \simeq {1\over 2}\,{\mpl\over f}\,{\phi_{50}\over f}
\la {\cal O}(0.1) ;\qquad
x_{50}^\prime \simeq - {1\over 2}\,\left( {\mpl\over f}\right)^2.
\end{equation}
Using the COBE DMR normalization, we can
relate $\Lambda$ to $f/\mpl$:
\begin{equation}
\Lambda /\mpl = 6.7\times 10^{-4} \sqrt{\mpl\over f}
\exp (-25\mpl^2/16\pi f^2).
\end{equation}
Further, we can solve for $T/S$, $\alpha_T$, and $\alpha_S$:
\begin{equation}
{T\over S} \simeq 0.07 \left( {\mpl\over f}\right)^2
\left({\phi_{50} \over f}\right)^2 \la {\cal O}(0.1) ;
\end{equation}
\begin{equation}
\alpha_T = {1\over 16\pi}\,{q_{50}\over q_{50}-1} \left(
{1\over 4} {\mpl^2\over f^2}{\phi_{50}^2\over f^2} \right)
\approx  {1\over 64\pi}
\left({\mpl\over f}\right)^2\left({\phi_{50}\over f}\right)^2\ll 0.1;
\end{equation}
\begin{equation}
\alpha_S = {1\over 16\pi}\,{q_{50}\over q_{50}-1} \left(
{1\over 4}{\mpl^2\over f^2}{\phi_{50}^2\over f^2} + {\mpl^2\over
f^2}\right) \approx {1\over 16\pi}\left({\mpl\over f}\right)^2;
\end{equation}
\begin{equation}
q_{50} = 64\pi \left({f\over \mpl}\right)^2\left(
{f\over \phi_{50}}\right)^2  \gg 1.
\end{equation}

Regime (ii) provides the exception
to the rule that $\alpha_S\approx\alpha_T$ and large
$\alpha_S$ implies large $T/S$.  For example, taking
$f=\mpl /2$, we find:
\begin{equation}
\phi_{50}/f \sim 0.06; \qquad x_{50} \sim 0.06; \qquad
x_{50}^\prime = - 2; \qquad q_{50} \sim 10^4;
\end{equation}
\begin{equation}
\alpha_T \sim 10^{-4};\qquad \alpha_S \sim 0.08;\qquad T/S \sim 10^{-3}.
\end{equation}
The gravitational-wave perturbations are very nearly scale
invariant, while the density perturbations deviate
significantly from scale invariance.  I note that
regime (ii), i.e., $f \la \mpl$, occupies only a tiny fraction of parameter
space because $f$ must
be greater than about $\mpl /3$ to achieve sufficient
inflation; further, regime (ii) is ``fine tuned'' and
``unnatural'' in the sense that the required value of $\Lambda$ is
exponentially sensitive to the value of $f/\mpl$.

Finally, I note that the results for regime (ii)
apply to any inflationary model whose Taylor expansion
in the inflationary region is similar; e.g., $V(\phi )=
-m^2\phi^2 + \lambda\phi^4$, which was originally analyzed
in Ref.~\cite{st}.

\subsubsection{Lessons}

To summarize the general features of our results.
In all examples the deviations from scale invariance
enhance perturbations on large scales.  The only
potentials that have significant deviations from
scale invariance are either very steep or have rapidly
changing steepness.  In the former case, both the
scalar and tensor perturbations are tilted by a
similar amount; in the latter case, only the scalar
perturbations are tilted.

For ``steep'' potentials,
the expansion rate is ``slow,'' i.e., $q_{50}$ close to unity,
the gravity-wave contribution to the CBR quadrupole anisotropy
becomes comparable to, or greater than, that of density
perturbations, and both scalar and tensor
perturbations are tilted significantly.
For flat potentials, i.e., small $x_{50}$,
the expansion rate is ``fast,'' i.e., $q_{50} \gg 1$,
the gravity-wave contribution to the CBR quadrupole
is much smaller than that of density perturbations, and unless
the steepness of the potential changes significantly,
large $x_{50}^\prime$, both spectra very nearly scale invariant;
if the steepness of the potential changes rapidly,
the spectrum of scalar perturbations can be tilted significantly.
The models that permit significant deviations from scale
invariance involve exponential or low-order polynomial
potentials; the former by virtue of their steepness, the latter
by virtue of the rapid variation of their steepness.
Exponential potentials are of interest because they arise
in extended inflation models; potentials with rapidly
steepness include $V(\phi ) =-m^2\phi^2+\lambda\phi^4$
or $\Lambda^4[1+\cos (\phi /f)]$.

Finally, to illustrate how observational data could used
to determine the properties of the inflationary potential
and test the consistency of the inflationary hypothesis,
suppose observations determined the following:
\begin{equation}
(\Delta T)_Q  \simeq 16\mu{\rm K};
\qquad T/S = 0.24; \qquad n = 0.9;
\end{equation}
that is, the COBE DMR quadrupole anisotropy, a four to one ratio
of scalar to tensor contribution to the CBR quadrupole,
and spectral index of 0.9 for the scalar perturbations.
{}From $T/S$, we determine the steepness of the potential:
$x_{50} \simeq 0.94$.  From the steepness and the
quadrupole anisotropy the value of the potential:
$V_{50}^{1/4}\simeq 2.4\times 10^{16}\GeV$.  From
the spectral index the change in steepness:
$x_{50}^\prime \simeq -0.81/\mpl$.
These data can also be expressed in terms of the
value of the potential and its first two derivatives:
\begin{equation}
V_{50} = 1.4\times 10^{-11}\mpl^4;\qquad
V_{50}^\prime = 1.5\times 10^{-11}\mpl^3;\qquad
V_{50}^{\prime\prime} = 1.0\times 10^{-12}\mpl^2.
\end{equation}
Further, they the lead to the prediction:
$n_T = -0.035$, which, when ``measured,'' can be used
as a consistency check for inflation.

\section{Structure Formation After COBE}

Filling in the details of structure formation is one
of the pressing challenges of the standard cosmology.
In order to do so one must have the ``initial data''
for the structure formation problem:  the spectrum
of density perturbations and the quantity and composition
of matter in the Universe.  With initial data in hand
one can hope to carry out detailed numerical simulations
which can be compared to the observations.\footnote{Since
the fluctuations predicted by inflation and other
theories are only specified in a statistical sense
this comparison can only be done statistically;
in the case of inflation, the fluctuations are
gaussian and so all predictions can be specified in
terms of the power spectrum, $\langle |\delta_k|^2\rangle$.}
While neither the observational data nor the simulations are
perfect, the situation in both regards is improving
rapidly.  In particular, the discovery of CBR anisotropy
by the COBE DMR has provided the first direct evidence for
the existence of density perturbations and thereby opened the
door for their study.

Over the past decade or
so many cosmologists have come to believe that required
initial data trace to events that took place during the
earliest history of the Universe ($t\ll 10^{-2}\sec$).
Thus, the study of structure formation has
the potential to test theories
of the early Universe and the underlying particle physics.
Inflation leads to two limiting scenarios:
hot dark matter and cold dark matter, both
with scale-invariant density perturbations.

In the hot dark matter scenario
the streaming of neutrinos from regions of higher density
to lower density erases perturbations on small scales
($\la 13h^{-2}\Mpc$); therefore
structure forms from the the top down:  superclusters
must form first and fragment into smaller objects.  Therein
lies the fundamental problem:  Since we know that superclusters
are just forming today, galaxies form too late to be
consistent with the abundance of galaxies observed at red shifts
of unity or so \cite{hdm}.

Cold dark matter looks much more promising; cold dark matter
refers to dark matter particles that move very slowly, either
by virtue of their large mass (e.g., 10 GeV to 2 TeV neutralino) or the
fact that they were born cold ($10^{-5}\eV$ axion).  This
means that perturbations on small scales are not erased
and that structure forms from the bottom up.
Cold dark matter has been subject to intense
scrutiny over the past decade and has thus far
survived, albeit with a number
of scratches and bruises \cite{jpo}.  CDM models
will be the focus of this Section.

That is not to say that cold dark matter models are the only
promising possibilities.  There are scenarios where the density
perturbations arise due to topological (and nontopological)
defects such as strings \cite{cs},
global monopoles, and textures \cite{texture}
with hot or cold dark matter.  Scenarios have been discussed
where the density perturbations arise in a rather recent
phase transition (since decoupling!), due to new physics in
the neutrino sector \cite{latetime}.

Finally, perhaps the most interesting
alternative is Peebles' PIB model or what-you-see-is-what-you-get
model \cite{pib}.  In PIB $\Omega_0=\Omega_B
\sim 0.2$, $h\sim 0.8$, and
the density perturbations are isocurvature perturbations
(variations in the local baryon-to-photon ratio and not
the energy density).  PIB is not motivated by what
early Universe theorists would like, rather by ``what
we see'' (though it violates the
primordial nucleosynthesis bound by large factor since
$\Omega_Bh^2 \simeq 0.13 \gg 0.02$).
Remarkably, the scenario is still
viable, though measurements of CBR anisotropy on scales
of $1^\circ$ to $90^\circ$ are really putting it to the
test:  normalizing to the COBE $10^\circ$ measurement,
its predictions for the quadrupole are a factor of
two small, while its predictions of scales of about five
degrees exceed current upper limits \cite{knoxmst}.

\subsection{The Universe observed}

By now we know a lot---and a little---about the structure
that exists in the Universe today.  A resurgence of
interest in structure formation, brought about in part
by the very intriguing early Universe suggestions
for initial data, has resulted in an explosion of observations that
bear on the issue over the past decade.
They include red-shift surveys (large-angle,
pencil-beam and sparsely sampled surveys), the spectrum
and spatial variation of the CBR temperature,
peculiar-velocity measurements, QSO absorption line
systems, studies of clusters and superclusters,
determinations of the distribution and quantity of dark matter,
studies of galactic evolution, catalogues of millions of galaxies
on the sky, and on and on.

To place things in perspective, we know much about the
distribution of light (bright galaxies)---as opposed to mass
(which is what theorists like to discuss);  the largest
red-shift survey, the CfA$_2$ slices of the Universe,
contains only about 20,000 galaxies with median red shift
of about 0.02 \cite{cfa}; and the total number of red shifts measured
for all purposes is only about 50,000.  We have no definitive evidence as to
the epoch of galaxy formation, or how the neutral hydrogen
left between galaxies became ionized (if it weren't, it
we would not be able to see emission from distant
QSOs shortward of Lyman alpha, 1215\AA\ in the rest frame
of the QSO).  We probably
only know the mean density of galaxies to within 20\%; we have
no fair sample of clusters; and so on.

Let me briefly try to summarize some of the
data that can be used to test models of structure
formation.  Within the spirit of my broad brush description,
I will group the observations into three classes:
Small-scale, observations that probe the Universe on
scales less than order $30h^{-1}\Mpc$ or so; Intermediate-scale,
observations that probe the Universe on scales of $30h^{-1}\Mpc
-300h^{-1}\Mpc$ or so; and Large-scale, observations that
probe the Universe on the very largest scales accessible.\footnote{I
warn the reader that my nomenclature is not universal;
many refer to what I call intermediate scales as large scales.}

\begin{itemize}

\item {\bf Small-scale Structure} ($\lambda \la 30h^{-1}\Mpc$):
Our knowledge of these scales is the most extensive
and well developed, though largely restricted to the
distribution of bright galaxies like our own.
These are also the scales on
which astrophysical effects---star formation,
blast waves, and so on---are potentially
most important and poorly understood.  The Universe on these scales
is organized into galaxies and clusters, whose properties
have been studied and quantified; for galaxies, number
density and morphology---i.e., spiral, elliptical,
etc.---rotation curves, and so on; and for clusters, number density,
velocity dispersions, richness class, and so on.
Both galaxies and clusters cluster, with
measured two-point correlation functions, $\xi_{\rm gg}(r)
\simeq (r/5h^{-1}\Mpc )^{-1.8}$ and $\xi_{\rm cc}(r)
\simeq (r/25h^{-1}\Mpc )^{-1.8}$, though the cluster
correlation function is less well known and depends
upon cluster richness \cite{bahcall}.  At some level we know the
distribution of dark matter:  spiral galaxies have
large halos with unknown spatial extent and
the bulk of the mass in clusters is dark \cite{dm}.
We also know the pairwise galaxy
velocity dispersion (line-of-sight velocity
dispersion), $\langle (v_1 -v_2)^2\rangle^{1/2}|_{10\Mpc}
\simeq 300-400\kms$ \cite{pairwise}.   (As discussed
earlier, the peculiar motions of galaxies
depend upon the amplitude of density perturbations
and the amount of matter in the Universe, and thus are indicative
of such.)  On scales less than about $8h^{-1}\Mpc$ the Universe is nonlinear:
specifically, the {\it rms} fluctuation in the number
density of bright galaxies measured
in a sphere of radius of $8h^{-1}\Mpc$ is unity.

\item {\bf Intermediate-scale Structure} ($30h^{-1}\Mpc
-300h^{-1}\Mpc$):  These are the scales on which our
knowledge is the most fragmentary and often more
qualitative than quantitative.\footnote{I often
call these the {\it NY Times} scales, as new observations and
their extravagant interpretation are reported there almost
weekly!}  Observations include the the voids and ``Great Wall''
seen in the CfA$_2$ red shift survey; the reoccurring walls
seen in the pencil-beam survey of Broadhurst et al. \cite{periodic};
the angular-correlation function of galaxies $w(\theta )$,
which is related to $\xi_{gg}(r)$, measured by Efstathiou
et al. \cite{apm} in the APM catalogue of 2 million galaxies on
the sky (effective depth of $400h^{-1}\Mpc$); the peculiar
velocities of galaxies measured by the Seven Samurai and
others \cite{peculiar}, about $400\kms$ on the
scale of $50h^{-1}\Mpc$; Great Attractors, and on and on.
{}From red-shift surveys like the CfA$_2$ slices of the Universe,
the IRAS $1.2\,$Jy survey of infrared-selected galaxies
\cite{fisheretal}, and the APM-Stromlo 1 in 20 red-shift survey
\cite{apmstromlo}, the fluctuations in the galaxy number density have
been measured on scales out to a few hundred Mpc; see Fig.~11a.
By the year 2000 the Sloan Digital Sky Survey \cite{SDSS} will produce
a ``Map of the Universe,'' from the
red shifts of a million galaxies (mean red shift of about
0.15 and survey depth of $500h^{-1}\Mpc$).
With the exception of the peculiar-velocity measurements
all these observations probe the distribution of light not mass.
CBR anisotropy measurements on angular scales of a few degrees
down to a few arcminutes also have the potential to
probe the distribution of matter on
these scales, as the CBR anisotropy on a given angular scale
is related to the fluctuations in the mass density
on a ranges of length scales around the characteristic
length scale that subtends that angular size on the
last scattering surface:  $\lambda \sim 100h^{-1}\Mpc
(\theta /{\rm deg}) $.  Very sensitive experiments are being
done on these angular scales; with the important exception
of the COBE DMR detection, there are now only
upper limits, at the level of a few times $10^{-5}$; see
Fig.~3.  I believe that more detections are just around the corner!

\item {\bf Large-scale Structure} ($\ga 300h^{-1}\Mpc$):
These scales are probed primarily by CBR anisotropy, though
the Sloan Digital Sky Survey should provide some information
about the distribution of galaxies on these scales.
On angular scales much greater than
about $1^\circ$ the anisotropy arises due to the
fluctuations in the gravitational potential on the
last-scattering surface (Sachs-Wolfe effect), while
on small-angular scales the situation is more complicated
as the velocity of the matter, temperature fluctuations
intrinsic to the radiation, and the ionization history of
the Universe become important.  On large-angular scales
it is very simple to relate the CBR anisotropy to the
``virgin spectrum'' of density fluctuations.  It is these scales that were
probed by the COBE DMR detection, proving the first direct information about
the existence of the density inhomogeneity that seeded
structure formation.

\end{itemize}

\subsubsection{Normalization:  the great leap forward!}
Lacking a definite prediction for the overall normalization
for inflationary density perturbations, those who
study formation of structure have historically used data
on small-scales to normalize the spectrum of density
perturbations, typically on the scale of $8h^{-1}\Mpc$.  In
so doing it is useful to define
\begin{equation}
\sigma_8 \equiv \langle ( \delta M /M
)^2\rangle^{1/2}_{8h^{-1}\Mpc};
\end{equation}
which is the {\it rms} mass fluctuation in spheres
of radius $8h^{-1}\Mpc$.
The simplest (and most naive) procedure is to
assume that light faithfully traces mass, i.e., $\delta\rho /\rho
= \delta n_{\rm GAL} /n_{\rm GAL}$, and set $\sigma_8
= 1$ since the {\it rms} fluctuation in galaxy number
in spheres of radius $8h^{-1}\Mpc$ is unity; I
will refer to this minimal cold dark matter (MCDM).
I should remark that there is no a priori reason to
expect light to trace mass, except on the very largest
scales where only gravity is important.

Because this normalization leads to a galaxy pairwise
velocity dispersion that is about a factor of two too large,
the concept of ``biasing'' was introduced; namely, that
light is a biased tracer of mass \cite{kaiser}.  If light
doesn't trace mass, the simplest
{\it ansatz} is a linear factor between the two:
\begin{equation}
\delta n_{\rm GAL}/n_{\rm GAL} = b (\delta \rho /\rho ).
\end{equation}
Of course there is every reason to expect that
the real relationship is more complicated, $b=b(\lambda )$.
In biased CDM models (BCDM), $\sigma_8
= b^{-1}$.  In principle, the bias factor $b(\lambda )$ can be measured
on scales where there is information about both the distribution
of galaxies and of mass, cf.~Fig.~11.

\begin{figure}
\vspace{4.0in}
\caption[iras]{Summary of observational knowledge
of the power spectrum $|\delta_k|^2$ based upon the
IRAS $1.2\,$Jy red shift survey and CBR anisotropy
measurements (from \cite{fisheretal}).  ACME-HEMT
indicates the South Pole experiment that has detected
anisotropy that may or may not be intrinsic to the CBR.}
\end{figure}

Until the COBE DMR detection,
a bias factor to 1.5 to 2 was in vogue to resolve the
discrepancy in the galaxy pairwise velocity dispersion; $b\sim 1.5-2$ was
known as the {\it standard} CDM model.
Since the peculiar velocities of galaxies arise due
to the lumpy distribution of matter; larger $b$
implies a smoother mass distribution and thus
smaller peculiar velocities.  (Likewise, reducing
the matter content, or $\Omega_0$, can help.)
Unfortunately, the predictions of BCDM on intermediate-scales
could not account for the level of inhomogeneity seen---voids,
galaxy-galaxy angular-correlation function,
peculiar velocities, and so on---since
the mass distribution was smoother.
Thus, CDM was faulted for predicting too
little power on ``large scales'' (in my nomenclature,
intermediate scales).

[Another motivation for bias is the so-called $\Omega$
problem:  Why do the dynamical measurements
indicate $\Omega_0 =0.2$, if $\Omega_0$ is really
unity?  The biasing explanation is
that most of the mass in the Universe is
in low surface-brightness galaxies that are too faint
to see and that are less strongly clustered than the
bright galaxies.  Bright galaxies are more strongly
clustered and account for only 20\% of the mass density.]

The COBE DMR detection of CBR anisotropy changed the
situation overnight by providing a new, more direct
normalization of the density perturbations!  Assuming the
correctness of the result, we now have a measurement
of the inhomogeneity in the mass distribution on
large-scales---and at last a ``physics normalization.''
Remarkably, the COBE normalization
(with scale-invariant perturbations)
corresponds to the simplest CDM model:
$\sigma_8=1.2\pm 0.2$ \cite{whiteetal}, i.e., no
biasing.\footnote{For HDM the COBE DMR normalization
implies $\sigma_8 = 0.7$.   This drives another nail
in the coffin, as it implies that only about 1\%
of the material in the Universe is in nonlinear structures.}

The COBE DMR normalization
has changed the way we view inflation and structure formation:
Intermediate (and large) scales seem to be OK; the problem is
with small scales.  Addressing this problem
is the focus of the brief discussion of CDM models that follows.

\subsection {CDM models}

The initial data for structure formation include:
(i) spectrum of primeval density
perturbations---amplitude on a given scale (normalization) and spectral
index $n$; (ii) composition of the Universe---$\Omega_i$,
$i$ = baryons, cold dark matter, hot dark matter, vacuum
energy and so on; and (iii) Hubble constant which sets the time/length
scale for the Universe.  With these in hand one can compute
the spectrum of density perturbations at the equivalence epoch
and let gravity run its course.  Of course, astrophysics---cooling
of baryons, star formation, etc.---is important too, but more
difficult to model.  Progress here too is being made with
large $N$-body codes that include both gravity and hydrodynamics
for the baryons \cite{sph}.  The list of wanted
cosmological parameters for a numerical simulation
is:  $\sigma_8$ (in the simple biasing prescription
$\sigma_8=b^{-1}$), $n$, $\Omega_B$, $\Omega_{\rm other}$, and $h$.

What predictions does inflation make for these parameters?
The firmest is a flat Universe, in my notation $\Omega_0 =1.0$,
which implies nonbaryonic dark matter dominates.
As mentioned earlier, hot dark matter ($30\eV$ or so neutrinos)
was ruled out early on; and so the cold dark matter scenario
{\it appeared} to be the unique inflationary blueprint for
structure formation \cite{cdm}.  Let me explain; to get the
age of the Universe right we must have $h\sim 0.5$.  This fact together
with the primordial nucleosynthesis determination of $\Omega_B
h^2$ implies $\Omega_B \simeq 0.04 -0.10$.  In {\it most} inflationary
models the density perturbations are very nearly scale invariant,
implying $n=1$.  Finally, the variance in galaxy counts on
$8h^{-1}\Mpc$ suggests $\sigma_8 =1$.  This is the
minimal cold dark matter model (MCDM); it is certainly the
simplest CDM scenario, though it is no longer the unique CDM model.

Partly due to problems with MCDM, partly due to the improvement
in the observations that test models of structure formation, and
partly due to the passage of time we now realize that there are
other possibilities, some just as well motivated, some less
well motivated.   I will characterize the different models
by their values for the key cosmological parameters
for structure formation:
$\sigma_8$, $n$, $\Omega_{\rm other}$, $h$, and $\Omega_B$.

\begin{figure}
\vspace{4in}
\caption[MCDM]{MCDM vs. observation.}
\end{figure}

\subsubsection{MCDM}  This is the simplest and the
original CDM model; it is characterized
by $b=1$, $n=1$, $\Omega_{\rm other} =\Omega_{\rm cold} \simeq 0.9$,
$\Omega_B \sim 0.1$, and $h=0.5$.  It is consistent with the
COBE DMR data, which for $n=1$ imply $\sigma_8=1.2\pm 0.2$,
and intermediate-scale structure.  However, it has
too much power on small scales, quantified  by a galaxy-pairwise
velocity dispersion of about $1000\kms$ compared to the
observed $400\kms$.  A comparison of MCDM power spectrum
with the observations is shown in Fig.~12.

Since MCDM is the simplest and
most well motivated model perhaps inflationists
should sit tight and wait
for the data (or their interpretations) to change.  After all,
the disagreement is on small scales where the Universe is highly
nonlinear and astrophysics can play an important role.

\subsubsection{BCDM}  This is the CDM model with biasing,
imposed to solve the problem of too much power on small scales.
The parameters of this model are:  $b\sim 1.5-2$, $n=1$, $\Omega_{\rm
other} =\Omega_{\rm cold}\sim 0.9$, $\Omega_B\sim 0.1$, and $h=0.5$.
This model is disfavor for two reasons: (1) The COBE DMR
results imply $b=0.8\pm 0.2$; and (2) (apparent)
insufficient power on intermediate scales to account
for peculiar velocities, the galaxy-galaxy angular correlation
function, etc.  However, one should keep in mind
that the COBE DMR results are new and may still
change, and that our knowledge of intermediate scales
is the least secure.  Perhaps the truth is somewhere in between MCDM
and BCDM; both are well motivated.   Shifting the
MCDM power spectrum in Fig.~12 downward by a factor of $2-4$
corresponds to $\sigma_8 \sim 0.5-0.7$.

\subsubsection{Tilt}  Tilted CDM (TCDM) models are characterized
by:  $\sigma_8\sim 0.5$, $n\sim 0.8$, $\Omega_{\rm
other} =\Omega_{\rm cold}\sim 0.9$, $\Omega_B\sim 0.1$, and $h=0.5$
\cite{tilt}.
{}From the beginning it was realized that
the inflationary perturbations were not precisely scale-invariant,
typically with more power on large scales ($n<1$) \cite{st}, and so
tilted models too are well motivated.
Relative to scale-invariant perturbations ($n=1$)
the density perturbation in a tilted model is
\begin{equation}
\left({\delta \rho \over\rho}\right)
\ \  \propto\ \  \left({\delta\rho \over\rho}\right)_{n=1}\,
\lambda^{(1-n)/2}.
\end{equation}
The COBE DMR result provides a normalization on very
large scales, $\lambda \sim 10^4\Mpc$; relative
to MCDM, the density perturbations on intermediate scales, $\lambda\sim
300\Mpc$, are only a factor of about 1.4 smaller,
while on small scales, $\lambda \sim 10\Mpc$, they
are about a factor of 2 smaller; see Fig.~13a.

If tilt is the truth, two kinds of inflationary
potentials are singled out; exponential and low-order
polynomial potentials \cite{tiltpot}.  Further, for exponential potentials,
the contribution of gravity waves to the CBR anisotropy
on large-angular scales is significant, which lowers
the overall normalization of density perturbations further,
by a factor of $(1+T/S)^{1/2}$ \cite{davisetal}.

\begin{figure}
\vspace{7.75in}
\caption[tcdm]{(a) TCDM vs. observation; (b) MDM/$\Lambda$CDM
vs. observation.}
\end{figure}

\subsubsection{Best-fit models}  These models address
the problem of too much small-scale power by changing
the transfer function.  The models considered
thus far are:  cold dark matter
with a cosmological constant ($\Lambda$CDM),
$n=1$, $\Omega_B \sim 0.05$, $\Omega_{\rm cold}
\sim 0.15$, $\Omega_\Lambda \sim 0.8$, and $h\sim 0.8$ \cite{lcdm};
and mixed dark matter (MDM)---``the neutrino
cocktail---''$n=1$, $\Omega_B\sim 0.1$, $\Omega_{\rm cold}\sim
0.6$, $h=0.5$, and $\Omega_{\rm hot} \sim 0.3$, corresponding
to a $7\eV$ or so mass neutrino \cite{mdm}.

How is the transfer function changed?  It is simplest to
see in MDM; since part of the dark matter is in the form
of neutrinos which freestream out of density fluctuations
on small scales, perturbations on small
scales are depressed (see Fig.~9), just what
the doctor ordered.  In $\Lambda$CDM the story is a little
more complicated; the bend in the transfer function is set
by the scale that crosses inside the horizon at matter-radiation
equality, $k_{\rm EQ} \sim 0.5 (\Omega_{\rm matter}h^2)
\Mpc$, where $\Omega_{\rm matter}=\Omega_B+\Omega_{\rm cold}
\simeq 0.2$.  Relative to MCDM, $k_{\rm EQ}$ is
a factor of two smaller, shifting the spectrum to
smaller $k$ and decreasing power on small scales (see Fig.~9).

[The $\Lambda$CDM model, which I once called ``the best-fit Universe,''
has a number of other nice features.  It automatically solves
the $\Omega$ problem since 80\% of the energy density is
in vacuum energy which is uniformly distributed and thus
does not ``show up'' in dynamical measurements of the
mass density.  It allows one to accommodate the higher values
of the Hubble constant which are favored by many measurements.
Likewise, MDM also address the $\Omega$ problem:  there
is not enough phase space in galaxies for neutrinos to
account for halo masses; further, neutrinos probably
move to fast to be captured even in clusters.  This would
explain why dynamical measurements of $\Omega_0$ based
on galactic rotation curves or cluster virial masses
do not lead to values of $\Omega_0$ close to unity.]

Because both $\Lambda$CDM and MDM have less power on
small scales than MCDM, when normalized to the
COBE DMR result, they are a much better fit to the
small scale data, and on large scales they are very
similar to MCDM.  They fit the present data very
well; see Fig.~13b.  One should, however, recall the
words of Francis Crick (of DNA fame); loosely quoted,
\begin{quotation}
\noindent A theory that agrees with all of the data at any
given time is necessarily wrong, since
at any given time some of the data are incorrect.
\end{quotation}

The weak point of the ``best-fit models'' is motivation;
however, let me try to make the best case for each.  $\Lambda$CDM:  A
cosmological constant that contributes an energy
density of about $10^{-46}\GeV^4$ would be very surprising.
Since there is no physical mechanism known that
explains why the present vacuum energy isn't of
order $\mpl^4$ (perhaps with the help
of supersymmetry only of order $G_F^{-2}$), one cannot
rigorously say that $\rho_{\rm vac} \simeq
10^{-46}\GeV^4$ is be fine-tuned in the technical sense.
MDM:  Cold dark matter has
so many good features that it must be {\it part}
of the truth; neutrinos exist---in three varieties---and
the see-saw mechanism suggests nondegenerate
masses in ``the $\eV$
range'' (meaning $10^{-6}\eV$ to tens of $\eV$),
it could well be that one of the three neutrinos has
a mass of order $10\eV$.

Certainly neither case for motivation is strong.
Why is the cosmological constant just today becoming
dynamically significant (recall, $\rho_{\rm vac}/\rho_{\rm matter}
\propto R^3$)?  And if history is any guide, cosmologists
beginning with Einstein have too often invoked a cosmological
constant to solve their problems.  For MDM, one has to
posit two kinds of nonbaryonic dark matter that each
contribute comparably to the energy density of the
Universe.  If nonbaryonic dark matter exists it is
already puzzling that baryons and dark matter
each contribute similar amounts to the mass density \cite{carr}.

\subsection{The scorecard and future}

Whereas cosmologists used to talk about ``the CDM
model'' and its ``uniqueness'' (words that were once
used to describe the superstring),
there is now a menu of CDM models.  How do they stand,
and which measurements can discriminate between them?
In discussing their models, it is often said that
theorists have many hands; on the one hand, on the
other hand, on the other other hand and so on.
Let me try my hand at it.

Occam's razor points to the simplest model, MCDM.
Moreover, it was vindicated by COBE
and only differs from the observational data by a
factor of two or so on small scales where complicated
astrophysics can be very important.  Perhaps theorists
should sit tight and wait.  On the other hand (here we
go), biasing at some level is likely to be a fact
of life, arguing for BCDM; BCDM resolves the
small-scale problems of CDM, but COBE indicates
that $b\sim 1$.  Maybe the truth is somewhere in between
MCDM and BCDM; the COBE normalization could come
down a bit, making $b\sim 1.3$ or so viable.

On the other other hand, deviation from scale-invariance
was in the cards from the beginning, and
so TCDM is well motivated too.  Moreover,
the tilt required points to a smaller class of
inflationary models, exponential potentials and
low-order polynomial potentials, which can be
discriminated between by the size of their tensor perturbations.

On my final hands are the best-fit models, $\Lambda$CDM
and MDM.  They are not as well motivated, but agree
better with the data at hand.  Of the two, my first
final hand has to go to MDM, and my second
final hand to $\Lambda$CDM.

There will be a variety of observations that can
used to discriminate between the different CDM models;
I will focus on CBR measurements
on the $0.5^\circ - 2^\circ$ scale, as there are
several experiments with the sensitivity to
probe CDM models which will be announcing results soon
\cite{fewdegree}.  These experiments add roughly
another order of magnitude to the range of scales
probed by CBR anisotropy (recall, COBE probes
$10^\circ$ to $90^\circ$).  For reference, the MCDM prediction
for this angular scale is $\delta T/T \sim 1-2\times
10^{-5}$; the current upper limits are just above this level!
Let me describe possible outcomes.

{\it Scenario 1:}  The upper limits become detections.
MCDM, $\Lambda$CDM, and MDM are in; TCDM and BCDM are out.

{\it Scenario 2:}  Detections are announced below
the $10^{-5}$ level.  BCDM and TCDM are in; the rest
are out.  If the detections are much below the $10^{-5}$
level, exponential potentials are strongly favored
as in these tilted models much of the COBE signal
is due to tensor perturbations whose contribution
to CBR anisotropy falls dramatically around a few
degrees \cite{tl,davisetal,bondetal}.

Finally, let me mention a very different test,
the value of the Hubble constant.
Suppose all parties agree on the currently popular value
$h=0.8$; all CDM models except $\Lambda$CDM fall by the
way side, based on the age of the Universe.
Conversely, suppose that evidence for $h=0.5$
becomes overwhelming; $\Lambda$CDM is out.

\section{Concluding Remarks}

Inflation is an extremely attractive cosmological
paradigm; in spite---no, because---of its beauty
it must be put to the ultimate test:  confrontation with
observation.  In testing inflation one must focus
on its robust predictions:  In order of robustness,
spatially flat Universe ($\Omega_0 =1$); very nearly
scale-invariant spectrum of density perturbations;
and nearly scale-invariant spectrum of
gravitational waves.  In addition,
for first-order inflation there should be a ``spike''
in the stochastic background of gravitational waves of
very significant energy density,
$\Omega_{\rm GW} \sim 10^{-9}$ or so, at a frequency
$f\sim 10^4\,{\rm Hz}({\cal M}/10^{10}\GeV )$ \cite{foigw}.  It is
also possible that fluctuations in other fields
lead to primeval magnetic fields \cite{mag} or isocurvature
perturbations (e.g., in axions or baryons) \cite{isocurv}.

There are a variety of means of testing these
predictions.  For example, there are kinematic
and dynamic techniques for measuring $\Omega_0$.
The density perturbations lead to temperature
fluctuations in the CBR.  The tensor (gravity-wave)
perturbations also lead to CBR anisotropies, or
may be detected directly by the next generation
of gravity wave detectors, Laser Interferometric
Gravity-wave Observatories (LIGOs) \cite{ligo}.

{}From the ``primary predictions,'' a series of secondary
predictions follow.  For example, since primordial
nucleosynthesis restricts $\Omega_B\la 0.1$, nonbaryonic
dark matter is a necessity, and a host of experiments
are under way to search for nonbaryonic dark matter
\cite{searchdm}.  In a flat,
matter-dominated Universe $H_0t_0 = {2\over 3}$ or
$t_0 \simeq 6.5h^{-1}\Gyr$, which implies that $h$
must be greater than 0.65 to ensure that the Universe
is older than 10 Gyr (the absolute minimum age that
is consistent with other measures of the age of the Universe).
The spectrum of density perturbations, together with
the matter content, provide the initial data for the
structure formation problem, leading to another test.

In the near term I believe that structure formation will provide
the most powerful test of inflation and probe of inflationary models.
On balance, the inflation-inspired CDM models are
doing quite well so far compared to
the alternatives:  Texture and cosmic-string models
required a high level of biasing ($b\sim 3$) to be compatible
with COBE DMR results; and PIB not only strongly violates
the primordial nucleosynthesis constraint to $\Omega_B$
but also seems to be inconsistent with CBR anisotropy
bounds on the scale of five degrees \cite{knoxmst}.
Great efforts are being made to further test the CDM scenarios,
and involve many different techniques, CBR anisotropy,
red-shift surveys, peculiar velocity measurements, and
so on.  These observations not only have the power to
falsify CDM, but could also reveal much
about the inflationary potential:  the value of the
potential, its steepness, and the change in steepness,
which in turn can used to learn about the underlying model.
For example, suppose that density perturbations do deviate
significantly from scale invariance, then two classes of
models are ruled out---chaotic and new inflation---and
two types of models are ruled in---exponential potentials
(as found in extended inflation) or low-order
polynomial potentials (as found in natural inflation).
The ratio of tensor to scalar perturbations can
further narrow the field:  large tensor contribution
to the CBR quadrupole points to exponential potentials and small-tensor
contribution points to low-order polynomial potentials.

The moment of truth for inflation may be near!

\bigskip\bigskip\bigskip
\noindent  This work was supported in part by the DOE
(at Chicago and Fermilab) and by the NASA through
grant NAGW-2381 (at Fermilab).


\begin{thebibliography} {distance1}

\bibitem{standard}  For a textbook treatment of the
standard cosmology see e.g., S.~Weinberg, {\it Gravitation
and Cosmology} (Wiley, NY, 1972); E.W.~Kolb and M.S.~Turner,
{\it The Early Universe} (Addison-Wesley, Redwood City, CA, 1990).

\bibitem{h50} A.~Sandage, {\it Physica Scripta} {\bf T43}, 22 (1992).

\bibitem{mould} J.~Mould et al., {\it Astrophys. J.}, in press (1993).

\bibitem{distance1}  See e.g., M.~Rowan-Robinson,
{\it The Cosmological Distance Ladder} (Freeman, San Francisco, 1985).

\bibitem{distance2}  M.~Fukugita, C.J.~Hogan, and
P.J.E.~Peebles, {\it Nature}, in press (1993).

\bibitem{FIRAS} J.~Mather et al., {\it Astrophys. J.}, in press (1993).

\bibitem{dnsnature} P.J.E.~Peebles, D.N.~Schramm,
E.~Turner, and R.~Kron, {\it Nature} {\bf 352}, 769 (1991).

\bibitem{COBRA} H.~Gush, M.~Halpern, and E.H.~Wishnow,
{\it Phys. Rev. Lett.} {\bf 65}, 537 (1990).

\bibitem{dipole}  G.F.~Smoot et al., {\it Astrophys. J.}
{\bf 396}, L1 (1992); D.J.~Fixsen et al., {\it ibid}, in press (1993).

\bibitem{yearly} G.F.~Smoot, in {\it First Course in Current
Topics in Astrofundamental Physics}, eds. N.~Sanchez and
A.~Zichichi (World Scientific, Singapore, 1992), p. 192.

\bibitem{DMR} G.F.~Smoot et al., {\it Astrophys. J.} {\bf 396}, L1 (1992);
E.L.~Wright, {\it ibid} {\bf 396}, L3 (1992).

\bibitem{davisetal} R.~Davis et al., {\it Phys. Rev. Lett.}
{\bf 69}, 1856 (1992).

\bibitem{wright} E.L.~Wright, {\it ibid} {\bf 396}, L3 (1992).

\bibitem{meyeretal} S.S.~Meyer, E.S.~Cheng, and L.A.~Page,
{\it Astrophys. J.} {\bf 371}, L1 (1991); K.~Ganga, S.S.~Meyer,
E.S.~Cheng, and L.A.~Page, {\it Astrophys. J.}, in press (1993).

\bibitem{walkeretal} T.P.~Walker et al., {\it Astrophys. J.}
{\bf 376}, 51 (1991).

\bibitem{tau} E.W.~Kolb et al., {\it Phys. Rev. Lett.}
{\bf 67}, 533 (1991).

\bibitem{dm} For recent reviews of dark matter see e.g.,
M.S.~Turner, {\it Physica Scripta} {\bf T36}, 167 (1991);
P.J.E.~Peebles, {\it Nature} {\bf 321}, 27 (1986);
V.~Trimble, {\it Ann. Rev. Astron. Astrophys.} {\bf 25}, 425 (1987);
J.~Kormendy and G.~Knapp, {\it Dark Matter in the Universe}
(Reidel, Dordrecht, 1989); K.~Ashman, {\it Proc. Astron. Soc. Pac.}
{\bf 104}, 1109 (1992); S.~Faber and J.~Gallagher, {\it Ann. Rev.
Astron. Astrophys.} {\bf 17}, 135 (1979).

\bibitem{interlopers}  S.~Faber and J.~Gallagher, {\it Ann. Rev.
Astron. Astrophys.} {\bf 17}, 135 (1979).

\bibitem{smallgroup} J.S.~Mulchaey, D.S.~Davis, R.F.~Mushotzky, and
D.~Burstein, {\it Astrophys. J.}, in press (1993).

\bibitem{irasomega} M.~Rowan-Robinson et al.,
{\it Mon. Not. R. astr. Soc.} {\bf 247}, 1 (1990);
N.~Kaiser et al., {\it ibid} {\bf 252}, 1 (1991);
M.~Strauss et al., {\it Astrophys. J.} {\bf 385}, 444 (1992).

\bibitem{potent}  E.~Bertschinger and A.~Dekel, {\it Astrophys. J.}
{\bf 336}, L5 (1989); A.~Dekel et al., {\it Astrophys. J.}, in press
(1993); M.~Strauss et al., {\it ibid} {\bf 397}, 395 (1992).

\bibitem{Sandage} A.~Sandage, {\it Astrophys. J.}
{\bf 133}, 355 (1961); {\it Physica Scripta} {\bf T43},
7 (1992); Refs.~[1].

\bibitem{lohspillar}  E.~Loh and E.~Spillar, {\it Astrophys. J.}
{\bf 307}, L1 (1986); M.~Fukugita et al., {\it ibid}
{\bf 361}, L1 (1990).

\bibitem{baryo} See e.g., E.W.~Kolb and M.S.~Turner,
{\it Ann. Rev. Nucl. Part. Sci.} {\bf 33}, 645 (1983);
A.~Dolgov, {\it Phys. Repts.}, in press (1993);
A.~Cohen, D.~Kaplan, and A.~Nelson, {\it Ann. Rev. Nucl.
Part. Sci.}, in press (1993).

\bibitem{Weinberg} S.~Weinberg, {\it Gravitation
and Cosmology} (Wiley, NY, 1972).

\bibitem{sf}  For a more complete pedagogical discussion
of structure formation see e.g., Refs.~[1]; P.J.E.~Peebles,
{\it The Large-scale Structure of the Universe}
(Princeton Univ. Press, Princeton, 1980); G.~Efstathiou,
in {\it The Physics of the Early Universe}, eds.~J.A.~Peacock,
A.F.~Heavens, and A.T.~Davies (Adam-Higler, Bristol, 1990).

\bibitem{dt/t} For a pedagogical discussion of CBR anisotropy
see e.g., G.~Efstathiou,
in {\it The Physics of the Early Universe}, eds.~J.A.~Peacock,
A.F.~Heavens, and A.T.~Davies (Adam-Higler, Bristol, 1990).
Also see, J.R.~Bond and G.~Efstathiou, {\it Mon. Not. R.
astr. Soc.} {\bf 226}, 655 (1987); J.R.~Bond et al.,
{\it Phys. Rev. Lett.} {\bf 66}, 2179 (1991).

\bibitem{SW} R.K.~Sachs and A.M.~Wolfe, {\it Astrophys. J.}
{\bf 147}, 73 (1967).

\bibitem{pib} P.J.E.~Peebles, {\it Nature} {\bf 327}, 210 (1987);
{\it Astrophys. J.} {\bf 315}, L73 (1987); R.~Cen, J.P.~Ostriker,
and P.J.E.~Peebles, {\it ibid}, in press (1993).

\bibitem{cs}  See e.g., A.~Vilenkin, {\it Phys. Repts.}
{\bf 121}, 263 (1985); A.~Albrecht and A.~Stebbins, {\it Phys.,
Rev. Lett.} {\bf 69}, 2615 (1992); D.~Bennett, A.~Stebbins,
and F.~Bouchet, {\it Astrophys. J.} {\bf 399}, L5 (1992).

\bibitem{texture}  See e.g., N.~Turok, {\it Phys. Rev.
Lett.} {\bf 63}, 2652 (1989); A.~Gooding, D.~Spergel, and N.
Turok, {\it Astrophys. J.} {\bf 372}, L5 (1991).

\bibitem{latetime}  J.~Fry, C.T.~Hill, and D.N.~Schramm,
{\it Comments on Nucl. Part. Phys.} {\bf 19}, 25 (1989);
A.~Gupta et al., {\it Phys. Rev. D} {\bf 45}, 441 (1992).

\bibitem{other} See e.g., C.W.~Misner, {\it Astrophys. J.}
{\bf 151}, 431 (1968); R.~Penrose, in {\it General Relativity:
An Einstein Centenary Survey}, eds. S.W.~Hawking and
W.~Israel (Cambridge Univ. Press, Cambridge, 1979);
R.H.~Dicke and P.J.E.~Peebles, {\it ibid}.

\bibitem{monopole}  J.~Preskill, {\it Ann. Rev. Nucl. Part.
Sci.} {\bf 34}, 461 (1984).

\bibitem{collins}  C.B.~Collins and S.W.~Hawking,
{\it Astrophys. J.} {\bf 180}, 317 (1973).

\bibitem{guth}  A.H.~Guth, {\it Phys. Rev. D} {\bf 23}, 347 (1981).

\bibitem{hu} Y.~Hu, M.S.~Turner, and E.J.~Weinberg, {\it Phys.
Rev. Lett.}, in press (1993).

\bibitem{coleman}  S.~Coleman, {\it Phys. Rev. D} {\bf 15}, 2929
(1977); S.~Coleman and R.~De Luccia, {\it ibid} {\bf 21},
3305 (1980).

\bibitem{bubblerh} R.~Watkins and L.~Widrow, {\it Nucl. Phys. B}
{\bf 374}, 446 (1992); S.W.~Hawking, J.~Stewart, and I.~Moss,
{\it Phys. Rev. D} {\bf 26}, 2681 (1982).

\bibitem{linde1}  A.D.~Linde, {\it Phys. Lett. B} {\bf 108},
389 (1982).

\bibitem{as} A.~Albrecht and P.J.~Steinhardt, {\it Phys. Rev. Lett.}
{\bf 48}, 1220 (1982).

\bibitem{reheat} A.~Albrecht et al., {\it Phys. Rev. Lett.}
{\bf 48}, 1437 (1982); L.~Abbott and M.~Wise, {\it Phys. Lett. B}
{\bf 117}, 29 (1982); A.D.~Linde and A.~Dolgov, {\it ibid}
{\bf 116}, 329 (1982).

\bibitem{foi}  For a review of first-order inflation see e.g.,
E.W.~Kolb, {\it Physica Scripta} {\bf T36}, 199 (1991).

\bibitem{extended}  D.~La and P.J.~Steinhardt,
{\it Phys. Rev. Lett.} {\bf 62}, 376 (1989).

\bibitem{twofield} F.~Adams and K.~Freese, {\it Phys. Rev. D}
{\bf 43}, 353 (1991).

\bibitem{kst}  E.W.~Kolb, D.~Salopek, and M.S.~Turner, {\it Phys.
Rev. D} {\bf 42}, 3925 (1990).

\bibitem{bigbubbles}  E.J.~Weinberg, {\it Phys. Rev. D} {\bf 40}, 3950 (1989);
M.S.~Turner, E.J.~Weinberg, and L.~Widrow, {\it ibid}
{\bf 46}, 2384 (1992).

\bibitem{sv}  Q.~Shafi and A.~Vilenkin, {\it Phys. Rev. Lett.}
{\bf 52}, 691 (1984).

\bibitem{Pi} S.-Y.~Pi, {\it Phys. Rev. Lett.} {\bf 52},
1725 (1984).

\bibitem{ewinflation} L.~Knox and M.S.~Turner, {\it Phys. Rev.
Lett.} {\bf 70}, 371 (1993).

\bibitem{olive} K.A.~Olive, {\it Phys. Repts.} {\bf 190}, 307 (1990).

\bibitem{hrr} R.~Holman, P.~Ramond, and G.G.~Ross, {\it Phys. Lett. B}
{\bf 137}, 343 (1984).

\bibitem{natural} K.~Freese, J.~Frieman, and
A.~Olinto, {\it Phys. Rev. Lett.} {\bf 65}, 3233 (1990).

\bibitem{chaotic} A.D.~Linde, {\it Phys. Lett. B} {\bf 129}, 177 (1983).

\bibitem{witterich} Q.~Shafi and C.~Wetterich, {\it Phys. Lett. B}
{\bf 129}, 387 (1983); {\it ibid} {\bf 152}, 51 (1985).

\bibitem{R2} A.A.~Starobinski, {\it Phys. Lett. B} {\bf 91}, 99 (1980);
M.B.~Mijic, M.S.~Morris, and W.-M.~Suen, {\it Phys. Rev. D}
{\bf 34}, 2934 (1986).

\bibitem{foigw}  M.S.~Turner and F.~Wilczek, {\it Phys.
Rev. Lett.} {\bf 65}, 3080 (1990);
A.~Kosowsky, M.S.~Turner, and R.~Watkins {\it ibid} {\bf 69}, 2026 (1992).

\bibitem{solar} R.D.~Reasenberg et al., {\it Astrophys. J.}
{\bf 234}, L219 (1979).

\bibitem{TW} M.S.~Turner and L.~Widrow, {\it Phys. Rev. Lett.}
{\bf 57}, 2237 (1986).

\bibitem{nohair} R.M.~Wald, {\it Phys. Rev. D} {\bf 28}, 2118
(1983); L.~Jensen and J.~Stein-Schabes, {\it ibid} {\bf 34},
931 (1986).

\bibitem{js} L.~Jensen and J.~Stein-Schabes, {\it Phys.
Rev. D} {\bf 35}, 1146 (1987).

\bibitem{aastar}  A.A.~Starobinskii, {\it JETP Lett.}
{\bf 37}, 66 (1983).

\bibitem{frieman} J.A.~Frieman and M.S.~Turner,
{\it Phys. Rev. D} {\bf 30}, 265 (1984).

\bibitem{piran} D.~Goldwirth and T.~Piran, {\it Phys. Repts.} {\bf 214},
223 (1992).

\bibitem{lindeqm} See e.g., A.D.~Linde, {\it Inflation and
Quantum Cosmology} (Academic Press, San Diego, CA, 1990).

\bibitem{scalar}  A.H.~Guth and S.-Y.~Pi, {\it Phys. Rev. Lett.}
{\bf 49}, 1110 (1982); A.A.~Starobinskii, {\it Phys. Lett. B}
{\bf 117}, 175 (1982); S.W.~Hawking, {\it ibid} {\bf 115}, 295 (1982);
J.M.~Bardeen, P.J.~Steinhardt, and M.S.~Turner, {\it Phys. Rev. D}
{\bf 28}, 679 (1983).

\bibitem{tensor} V.A.~Rubakov, M.~Sazhin, and A.~Veryaskin,
{\it Phys. Lett. B} {\bf 115}, 189 (1982); R.~Fabbri and
M.~Pollock, {\it ibid} {\bf 125}, 445 (1983); L.~Abbott
and M.~Wise, {\it Nucl. Phys. B} {\bf 244}, 541 (1984);
B.~Allen, {\it Phys. Rev. D} {\bf 37}, 2078 (1988).

\bibitem{allslow}  At first sight, first-order inflation
might seem very different
from slow-rollover inflation, as reheating occurs through
the nucleation of percolation of true-vacuum bubbles.
However, such models can be recast as slow-rollover inflation by means
of a conformal transformation, and the analysis of
metric perturbations proceeds as in slow rollover inflation.
See e.g., E.W.~Kolb, D.~Salopek, and M.S.~Turner, {\it Phys.
Rev. D} {\bf 42}, 3925 (1990).

\bibitem{isocurv} See e.g., A.D.~Linde, {\it Phys. Lett. B}
{\bf 158}, 375 (1985); D.~Seckel and M.S.~Turner,
{\it Phys. Rev. D} {\bf 32}, 3178 (1985); M.S.~Turner,
A.~Cohen, and D.~Kaplan, {\it Phys. Lett. B} {\bf 216}, 20 (1989).

\bibitem{inflation} E.W.~Kolb and M.S.~Turner, {\it The
Early Universe} (Addison-Wesley, Redwood City, CA, 1990),
Ch.~8.

\bibitem{hz} E.R.~Harrison, {\it Phys. Rev. D} {\bf 1},
2726 (1970); Ya.B.~Zel'dovich, {\it Mon. Not. R. astr. Soc.}
{\bf 160}, 1p (1972).

\bibitem{st} P.J.~Steinhardt and M.S.~Turner, {\it Phys. Rev. D}
{\bf 29}, 2162 (1984).

\bibitem{turner}  The material presented in this Section is a summary
of work completed during this school and will be published
elsewhere, M.S.~Turner, {\it Phys. Rev. D}, in press (1993).

\bibitem{stat} J.M.~Bardeen et al., {\it Astrophys. J.}
{\bf 304}, 15 (1986).

\bibitem{ls} D.H.~Lyth and E.D.~Stewart, {\it Phys. Lett. B}
{\bf 274}, 168 (1992); E.D.~Stewart and D.H.~Lyth, {\it ibid},
in press (1993).

\bibitem{aw} L.~Abbott and M.~Wise, {\it Nucl. Phys. B}
{\bf 244}, 541 (1984).

\bibitem{white} M.~White, {\it Phys. Rev. D} {\bf 46}, 4198 (1992).

\bibitem{turner} M.S.~Turner, {\it Phys. Rev. D}, in press (1993).

\bibitem{tl} M.S.~Turner and J.E.~Lidsey, {\it Phys. Rev. D},
in press (1993).

\bibitem{powerlaw} L.~Abbott and M.~Wise, {\it Nucl. Phys. B}
{\bf 244}, 541 (1984); F. Lucchin and S. Mattarese, {\it Phys. Rev. D}
{\bf 32}, 1316 (1985); R.~Fabbri, F.~Lucchin, and S.~Mattarese,
{\it Phys. Lett. B} {\bf 166}, 49 (1986).

\bibitem{massive}  V.~Belinsky, L.~Grishchuk, I.~Khalatanikov,
and Ya.B.~Zel'dovich, {\it Phys. Lett. B} {\bf 155}, 232
(1985); L.~Jensen, unpublished (1985).

\bibitem{star} A.A.~Starobinskii, {\it Sov. Astron.} {\bf 11}, 133 (1985).

\bibitem{new} A.D.~Linde, {\it Phys. Lett. B}
{\bf 108}, 389 (1982); A.~Albrecht and P.J.~Steinhardt,
{\it Phys. Rev. Lett.} {\bf 48}, 1220 (1982).

\bibitem{hdm} S.D.M.~White, C.~Frenk, and M.~Davis,
{\it Astrophys. J.} {\bf 274}, L1 (1983); {\it ibid}
{\bf 287}, 1 (1983); J.~Centrella and A.~Melott, {\it Nature}
{\bf 305}, 196 (1982).

\bibitem{jpo} J.P.~Ostriker, {\it Ann. Rev. Astron. Astrophys.}
{\bf 31}, in press (1993).

\bibitem{knoxmst} L.~Knox and M.S.~Turner, {\it Astrophys. J.},
in press (1993).

\bibitem{cfa}  V.~De Lapparent, M.~Geller, and J.~Huchra,
{\it Astrophys. J.} {\bf 302}, L1 (1986); {\it ibid}
{\bf 332}, 44 (1988); M.~Geller and
J.~Huchra, {\it Science} {\bf 246}, 897 (1989).

\bibitem{bahcall} N.~Bahcall, {\it Ann. Rev. Astron. Astrophys.}
{\bf 26}, 631 (1988).

\bibitem{pairwise} M.~Davis et al., {\it Astrophys. J.}
{\bf 292}, 371 (1985).

\bibitem{periodic} T.~Broadhurst et al., {\it Nature}
{\bf 343}, 726 (1990).

\bibitem{apm} S.J.~Maddox et al., {\it Mon. Not. R. astr.
Soc.} {\bf 242}, 43p (1990).

\bibitem{peculiar} A.~Dressler et al., {\it Astrophys. J.}
{\bf 313}, L37 (1987); E.~Bertschinger et al., {\it ibid}
{\bf 364}, 370 (1990) and references therein.

\bibitem{fisheretal} K.~Fisher et al., {\it Astrophys. J.}
{\bf 389}, 188 (1992); also see, M.S.~Vogeley et al.,
{\it ibid} {\bf 391}, L5 (1992); W.~Saunders et al.,
{\it Nature} {\bf 349}, 42 (1991).

\bibitem{apmstromlo} J.~Loveday et al., {\it Astrophys. J.}
{\bf 390}, 338 (1992); {\it ibid} {\bf 400}, L43 (1992).

\bibitem{SDSS} The Sloan Digital Sky Survey is a collaboration
between The University of Chicago, Fermilab, Johns Hopkins
University, Princeton University and the Institute for Advanced
Study.

\bibitem{kaiser} N.~Kaiser, {\it Astrophys. J.} {\bf 284}, L9 (1984);
in {\it Inner Space/Outer Space}, eds. E.W.~Kolb et al. (Univ. of
Chicago Press, Chicago, 1986), p. 258.

\bibitem{whiteetal}  G.~Efstathiou, J.R.~Bond, and
S.D.M.~White, {\it Mon. Not. R. astr. Soc.} {\bf 258}, 1p (1992).

\bibitem{sph} R.Y.~Cen and J.P.~Ostriker, {\it Astrophys. J.}
{\bf 393}, 22 (1992); A.E.~Evrard, F.J.~Summers, and M.~Davis,
{\it Astrophys. J.}, in press (1993); E.~Bertschinger and J.~Gelb,
{\it Computers in Physics} {\bf 5}, 164 (1991); J.~Gelb and
E.~Bertschinger, {\it Astrophys. J.}, in press (1993);
N.~Katz, L.~Hernquist, and D.~Weinberg, {\it Astrophys. J.}
{\bf 399}, L109 (1992); G.~Evrard, {\it Mon. Not. R. astr. Soc.}
{\bf 235}, 911 (1988); C.~Park and J.R.~Gott {\it ibid} {\bf 249}, 288 (1991);
C.~Park, {\it ibid} {\bf 242}, 59p (1990);
C.~Frenk et al., {\it Astrophys. J.} {\bf 351}, 10 (1990).

\bibitem{cdm} For a synopsis of structure formation in
the CDM scenario see e.g., G.R.~Blumenthal et al., {\it Nature}
{\bf 311}, 517 (1984).

\bibitem{tilt}  See e.g., J.P.~Ostriker, {\it Ann. Rev. Astron.
Astrophys.} {\bf 31}, in press (1993);
F.~Adams et al., {\it Phys. Rev. D} {\bf 47}, 426
(1993); J.~Gelb et al., {\it Astrophys. J.} {\bf 403},
L5 (1993); R.~Cen and J.P.~Ostriker, {\it ibid}, in press (1993).

\bibitem{tiltpot} R.~Davis et al., {\it Phys. Rev. Lett.}
{\bf 69}, 1856 (1992); F.~Lucchin, S.~Mattarese, and
S.~Mollerach, {\it Astrophys. J.} {\bf 401}, L49 (1992); D.~Salopek,
{\it Phys. Rev. Lett.} {\bf 69}, 3602 (1992); A.~Liddle and D.~Lyth,
{\it Phys. Lett. B} {\bf 291}, 391 (1992); J.E.~Lidsey and
P.~Coles, Queen Mary College preprint (1992); A.~Dolgov and J.~Silk,
unpublished (1992); T.~Souradeep and V.~Sahni, {\it Mod. Phys. Lett.
A} {\bf 7}, 3541 (1992).

\bibitem{lcdm} M.S.~Turner, G.~Steigman, and L.~Krauss,
{\it Phys. Rev. Lett.} {\bf 52}, 2090 (1984); M.S.~Turner,
{\it Physica Scripta} {\bf T36}, 167 (1991); P.J.E.~Peebles,
{\it Astrophys. J.} {\bf 284}, 439 (1984); G.~Efstathiou et al.,
{\it Nature} {\bf 348}, 705 (1990).

\bibitem{mdm} Q.~Shafi and F.~Stecker,
{\it Phys. Rev. Lett.} {\bf 53}, 1292 (1984);
S.~Achilli, F.~Occhionero, and R.~Scaramella,
{\it Astrophys. J.} {\bf 299}, 577 (1985);
S.~Ikeuchi, C.~Norman, and Y.~Zahn,
{\it Astrophys. J.} {\bf 324}, 33 (1988);
A.~van Dalen and R.K.~Schaefer, {\it Astrophys. J.}
{\bf 398}, 33 (1992); M.~Davis, F.~Summers, and D.~Schlegel,
{\it Nature} {\bf 359}, 393 (1992); J.~Holtzman and J.A.~Primack,
{\it Astrophys. J.}, in press (1993); A.~Klypin et al.,
{\it Astrophys. J.}, in press (1993); D.~Pogosyan and
A.A.~Starobinsky, DAMTP/IOA/MRAO preprint (1993).

\bibitem{carr} M.S.~Turner and B.J.~Carr, {\it Mod. Phys. Lett. A}
{\bf 2}, 1 (1987).

\bibitem{fewdegree} T.~Gaier et al., {\it Astrophys. J.}
{\bf 398}, L1 (1992); D.C.~Alsop et al., {\it ibid}
{\bf 387}, 146 (1992); A.C.S.~Readhead et al., {\it ibid}
{\bf 346}, 566 (1989); P.~De Bernardis et al., {\it ibid}
{\bf 396}, L57 (1992); R.A.~Watson et al., {\it Nature}
{\bf 357}, 660 (1992); S.S.~Meyer, E.S.~Cheng, and L.A.~Page,
{\it Astrophys. J.} {\bf 371}, L1 (1991); J.O.~Gundersen et al.,
{\it ibid}, in press (1993); P.R.~Meinhold et al., {\it ibid},
in press (1993).

\bibitem{bondetal}  J.R.~Bond, R.~Crittenden, R.~Davis,
G.~Efstathiou, and P.J.~Steinhardt, {\it Phys. Rev. Lett.},
in press (1993).

\bibitem{mag} M.S.~Turner and L.M.~Widrow, {\it Phys. Rev. D}
{\bf 37}, 2743 (1988); B.~Ratra, {\it Astrophys. J.}
{\bf 391}, L1 (1992).

\bibitem{ligo} A.~Abramovici et al., {\it Science}
{\bf 256}, 325 (1992); K.S.~Thorne, in {\it 300 Years of
Gravitation}, eds. S.W.~Hawking and W.~Israel (Cambridge
Univ. Press, Cambridge, 1987), p. 330.

\bibitem{searchdm}  See e.g., J.R.~Primack, D.~Seckel, and B.~Sadoulet,
{\it Ann. Rev. Nucl. Part. Sci.} {\bf 38}, 751 (1988);
D.O.~Caldwell, {\it Mod. Phys. Lett. A} {\bf 5}, 1543 (1990);
P.F.~Smith and J.D.~Lewin, {\it Phys. Repts.} {\bf 187}, 203 (1990).

\end{thebibliography}
\end{document}